\documentclass[prd,showkeys,showpacs,preprintnumbers,byrevtex]{revtex4}
\usepackage{amsmath,amsfonts,amssymb,amscd,amsxtra,amsthm}
\usepackage{graphicx}
\usepackage{bm}
\usepackage{epstopdf}
\begin{document}
\preprint{KIAS-P12087}
\title{Electroproduction of $\Lambda(1520)$ off the nucleon target with the nucleon resonances}
%-------------------------------------------------
\author{Seung-il Nam}
\email[E-mail: ]{sinam@pknu.ac.kr; sinam@kias.re.kr}
\affiliation{School of Physics, Korea Institute for Advanced Study (KIAS), Seoul 130-722, Republic of Korea}
\affiliation{Department of Physics, Pukyong National University (PKNU), Busan 608-737, Republic of Korea}
%-------------------------------------------------
\date{\today}
\begin{abstract}
We investigate the unpolarized electroproduction of $\Lambda(1520)\equiv\Lambda^*$ off the nucleon target, using the effective Lagrangian method at the tree-level Born approximation with the nucleon-resonance contributions from $S_{11}(2090)$, $D_{13}(2080)$, and $D_{15}(2200)$. First, we compute the various physical quantities for the proton target case, such as the total and differential cross sections, $t$-momentum transfer distribution, and $K^-$ decay-angle ($\phi$) distribution.  It turns out that $D_{13}$ plays an important role to reproduce the electroproduction data properly. The numerical results for the $\phi$ distribution shows obvious different structures from that for the photoproduction, due to the enhancement of the kaon exchange by the longitudinal polarization of the virtual photon as expected. Numerically, we observe that the kaon-exchange contribution in the $t$ channel becomes about a half of that from the contact-term one that dominates the photoproduction of $\Lambda^*$. We also provide theoretical estimations for the $\Lambda^*$ electroproduction off the neutron target, showing that its production rate is saturated almost by the resonance contributions.  Finally, the contact-term dominance, which is the key ingredient for the $\Lambda^*$ electromagnetic productions, is briefly discussed. 
\end{abstract}
\maketitle
%--------------------------------------------------
\section{Introduction}
%--------------------------------------------------
Electromagnetic (EM) productions of the hadrons off the nucleon target have been very useful tools to study the nonperturbative quantum chromodynamics (QCD) in terms of the color-singlet degrees of freedom at a scale $\sim1$ GeV from the experimental and theoretical points of view. From those production processes, one can extract the fundamental information for the EM and strong interaction structures, the hadron mass spectra, and missing resonances searches for instance. Note that the EM production of hadrons have been studied energetically by the various experimental collaborations: LAMP2 at Daresbury~\cite{Barber:1980zv}, CLAS at Jefferson Laboratory~\cite{Barrow:2001ds}, LESP at SPring-8~\cite{Muramatsu:2009zp,Kohri:2009xe,Zhao:2010zzm}, CB-ELSA/TAPS at Bonn~\cite{Nanova:2008kr}, and so on. Along with those experimental endeavors, there have been abundant theoretical works for them as well, such as the photoproduction of $K\Lambda(1520)$~\cite{Nam:2005uq,Toki:2007ab,Nam:2009cv,Nam:2010au,Xie:2010yk,He:2012ud}, $\eta N$~\cite{Choi:2007gy}, $\pi\Delta(1231)$~\cite{Nam:2011np}, $K\Lambda(1116)$~\cite{Janssen:2001wk}, $K^*\Lambda(1116)$~\cite{Oh:2006hm,Kim:2011rm}, and $K^*\Sigma(1190)$~\cite{Kim:2012pz}, employing the tree-level Born approximation with the effective Lagrangian approach, and accumulated considerably important results. 

Among those theoretical efforts, it is worth mentioning the interesting results of our previous works on the $\Lambda(1520,3/2^-)\equiv\Lambda^*$ photoproduction~\cite{Nam:2005uq,Nam:2009cv,Nam:2010au}. In terms of the gauge invariance of the scattering amplitude, i.e. the Ward-Takahashi (WT) identity, it turned out that the contact-term contribution prevails over all other kinematic channels. This interesting behavior of the {\it contact-term dominance} can be also resulted in the large target asymmetry, saying that the production rate from the proton target is much larger than that for the neutron-target case, in which the contact-term contribution does not exists due to the electric-charge conservation: $\sigma_n\ll\sigma_p$. In 2009, the LEPS collaboration reported that the differential cross sections for the $\Lambda^*$ photoproduction off the proton as well as the deuteron targets. Interestingly enough, the production rates turned out to be similar to each other: $d\sigma_d/d\cos\theta\sim d\sigma_p/d\cos\theta$. Here, $\theta$ denotes the angle for the outgoing $K^+$ with respect to the incident photon in the center-of-mass (cm) system. Thus, this observation indicates that the contact-term dominance works qualitatively well, if we take into account a naive but reasonable assumption $\sigma_d\sim\sigma_p+\sigma_n$~\cite{Muramatsu:2009zp}. Note that there are theoretical supports for the contact-term dominance for $\gamma p \to K^+\Lambda^*$~\cite{Toki:2007ab,Zhao:2010zzm}.

The resonance contributions for the $\Lambda^*$ production is also important to be studied.  From  the photoproduction experiment by the LEPS collaboration~\cite{Kohri:2009xe}, it was observed that the peak of the differential cross section as a function of $E_\gamma$ varies its strength depending on the $\theta$ angle. This tendency may indicate a possible contribution from a nucleon resonance. Then, the two theoretical works, employing the effective approaches, suggested that the $D_{13}(2080)$ resonance, which is now split into two different resonant states $D_{13}(2120)$ and $D_{13}(1875)$~\cite{Nakamura:2010zzi}, can be the most possible candidate for it~\cite{Xie:2010yk,He:2012ud}. We also want to focus on the distinctive features between the electro and photoproductions of $\Lambda^*$. In the CLAS experiment for the $\Lambda^*$ electroproduction~\cite{Barrow:2001ds}, the $K^-$ decay-angle ($\phi$) distribution in the Gottfried-Jackson (GJ) frame was described mainly by a curve $\propto(1/3+\cos^2\phi)$, signaling that $\Lambda^*(S=1/2)$ decays into $K^-p$ mainly, in addition to small contributions from the other spin states. Here, $\Lambda^*(S)$ stand for that $\Lambda^*$ in its spin-$S$ states, and $\phi$ indicates the angle between the target nucleon and outgoing $K^-$ in the $\Lambda^*$ rest frame. In contrast, the $\phi$ distribution for the $\Lambda^*$ photoproduction shows the curve $\propto\sin^2\phi$, in which $\Lambda^*(S=3/2)$  decays into $K^-p$~\cite{Barber:1980zv,Muramatsu:2009zp}. In our previous work~\cite{Nam:2010au}, the $\phi$ distribution for the photoproduction was reproduced qualitatively well in terms of the contact-term dominance. This result also indicates that the contact-term contribution plays the role of the spin-$1$ meson exchange in addition to the $K^*$ exchange in the $t$ channel. In general, the most obvious difference in these two EM productions is that the existence of the longitudinal-polarization component of the virtual photon. Physically, the difference shown in the $\phi$ distribution between the EM productions can be understood by the $S_z=0$ (longitudinal) component of the polarization vector selects the spin-$0$ meson, i.e. kaon exchange in the $t$ channel. Hence, in the electroproduction, the possibility for $\Lambda^*(S=1/2)$ gets enhanced than that for $\Lambda^*(S=3/2)$, resulting in that the curve shape for the $\phi$ distribution becomes proportional to $\propto(1/3+\cos^2\phi)$.

Focusing on the two intriguing ingredients mentioned above, 1) the nucleon-resonance contributions and 2) the distinctive features in the electroproduction in comparison to the photoproduction,  in the present work, we want to investigate the elementary process for the electroproduction of $\Lambda(1520,3/2^-)\equiv\Lambda^*$ off the nucleon target, i.e. $\gamma^*N\to K\Lambda^*$. So far, we have had only one $\Lambda^*$-electroproduction experiment done by Barrow {\it et al.} of the CLAS collaboration~\cite{Barrow:2001ds}. Hence, we will closely explore those data theoretically in the present work. To this end, we make use of the effective Lagrangian approach at the tree-level Born approximation, closely following the theoretical framework in our previous  works~\cite{Nam:2005uq,Nam:2009cv,Nam:2010au}. Note that this very simple theoretical framework violates unitarity explicitly. This unitarity problem can be cured by considering intermediate scattering processes with the mesons and baryons, such as the $K$-matrix method~\cite{Shyam:2009za}. A typical calculation using the $K$-matrix method for the $S=0$ channel was done for the $\phi$ photoproduction, taking into account the intermediate $K\Lambda^*$ state in addition to the ground states, in Ref.~\cite{Ozaki:2009mj}. It turned out that the effects of the $K$-matrix unitarization is small, and the pomeron-exchange dominates both of the $K$-matrix and tree-level results.  Hence, considering the similarity within this kind of calculations, we would like to keep employing the tree-level calculation for the $\Lambda^*$ EM production processes, avoiding complexities in the calculations. It is also worth mentioning that, in Ref.~\cite{Davidson:1991xz}, the authors explored several effective unitarization methods within the effective Lagrangian approach, although the inclusion of the unitarity into the tree-level approximation is still imperfect in terms of the gauge invariance, relativity, and so on. The Rarita-Schwinger vector-spinor formalism is used for describing the spin-$3/2$ field for $\Lambda^*$ in a field theoretical manner~\cite{Rarita:1941mf,Nath:1971wp}. 

We also take into account the phenomenological form factors for each kinematic channels, following the prescription suggested in Refs.~\cite{Haberzettl:1998eq,Davidson:2001rk,Haberzettl:2006bn}, with a newly devised gauge-conserving term similar to that given in Ref.~\cite{Mart:2010ch}. To take into account the $Q^2$ dependence of the form factors, we employ the dipole and monopole EM form factors for the vertices, where the hadrons couple to the virtual photon. All the numerical calculations are performed in the $K$-$\Lambda^*$ cm system with the properly defined four momenta and polarization vectors for the relevant particles involved. The transverse polarization parameter $\varepsilon$ is chosen to be $0.5$ throughout the theoretical calculations, considering the experimentally given value $\varepsilon=(0.3\sim0.7)$~\cite{Barrow:2001ds}. In addition, taking into account the CLAS experiment, we choose the kinematic regions for the photon virtuality and $K$-$\Lambda^*$ cm energy  as $Q^2=(0.9\sim2.4)\,\mathrm{GeV}^2$ and $W=(1.95\sim2.65)$ GeV, respectively. In the present work, as a first step to investigate the nucleon-resonance effects for the scattering process, we take into account $S_{11}(2090)$, $D_{13}(2080)$, and $D_{15}(2200)$, since these resonances were found to be relevant in the $\Lambda^*$ photoproduction~\cite{Xie:2010yk,He:2012ud}. In order to study the $\phi$ distribution, we devise a simple parameterization for it, by separating the differential cross section into each $\Lambda^*$ spin states theoretically~\cite{Nam:2010au}.

Firstly, as for the theoretical results, we provide the total and differential cross sections, $\Lambda^*$ spin distribution, $\phi$ distribution, and $t$-momentum transfer distribution for the proton target case. It turns out that the present theoretical framework reproduces the data qualitatively well with the resonance contributions, which play a crucial role in the vicinity of $W\lesssim2.4$ GeV. At the same time, it turns out that the $D_{13}$ resonance enhances the production rate in the forward-scattering region. We also show numerically that the longitudinal-polarization of the virtual photon enhances the kaon-exchange contribution more as expected, than its effects on the contact-term one. This enhancement is also shown explicitly by that the numerical results for the $\phi$ distribution turns out to be considerably different from that for the photoproduction. Secondly, we apply the determined information for the resonances from the electroproduction to the photoproduction of $\Lambda^*$ to see their effects. By doing that, we observe that the resonance contributions are not so effective in comparison to that for the electroproduction, since the contact-term contribution dominates the photoproduction process even with the resonance. Finally, using all the ingredients obtained above, we compute the total and differential cross sections for the $\Lambda^*$ electroproduction off the neutron target. On top of the negligible $K^*$-exchange contribution, the production rate of the cross sections are saturated almost by the $D_{13}$ and $D_{15}$ contributions. From this, we can conclude that the $\Lambda^*$ EM productions off the neutron target are utmost useful production channels to investigate the nucleon-resonance contributions, since the production rate is dominated almost only by them, due to the absence of the {\it contact-term} contribution. Moreover, we make some brief comments on the contact-term dominance in the $\Lambda^*$ EM productions, considering all the observations mentioned above. 

The present work is organized as follows: In Section II, we briefly introduce the present theoretical framework, defining the effective interactions for the relevant Yukawa vertices, computing coupling strengths, writing down the invariant amplitudes, and so on. The numerical results and related discussions are given in Section III. Section IV is devoted for summary and future prospectives.  

%--------------------------------------------------
\section{Theoretical framework}
%--------------------------------------------------
%FIGURE>>>
\begin{figure}[t]
\includegraphics[width=14cm]{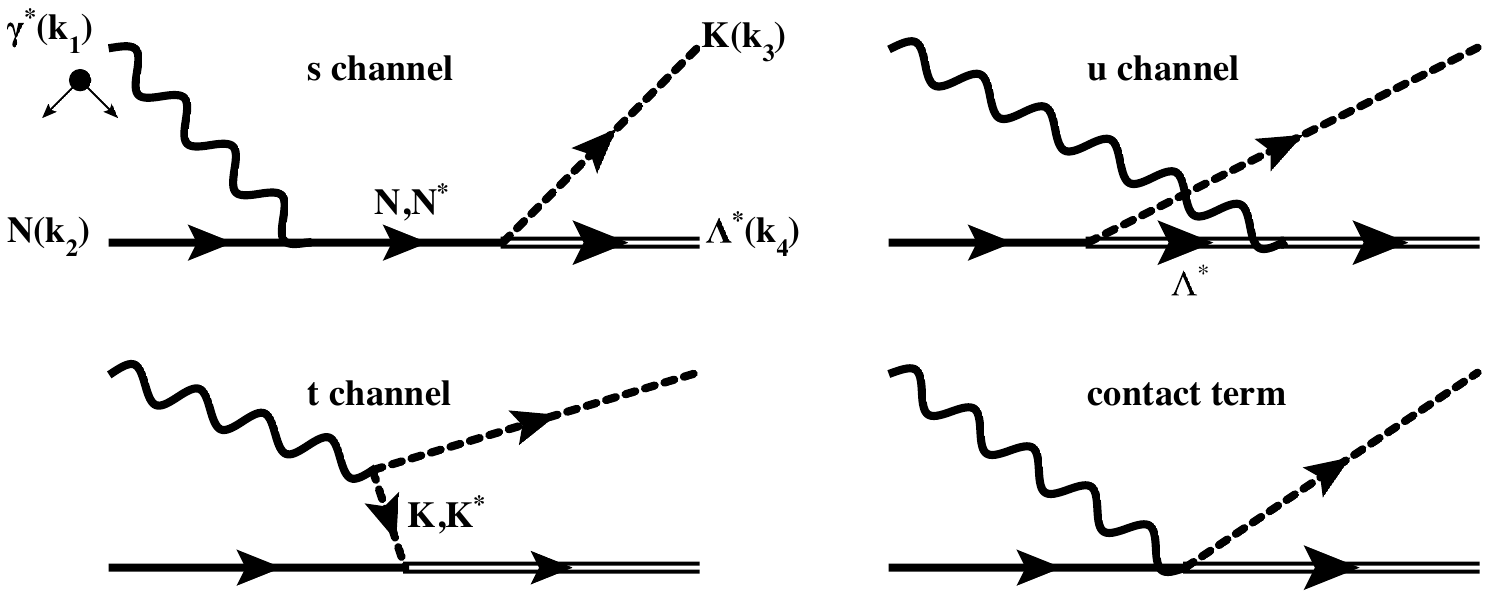}
\caption{Tree-level Feynman diagrams for $\gamma^*N\to K\Lambda^*$. Relevant momenta of the particles involved are defined in Eq.~(\ref{eq:MOMENTA}).}       
\label{FIG0}
\end{figure}
\begin{figure}[t]
\includegraphics[width=10cm]{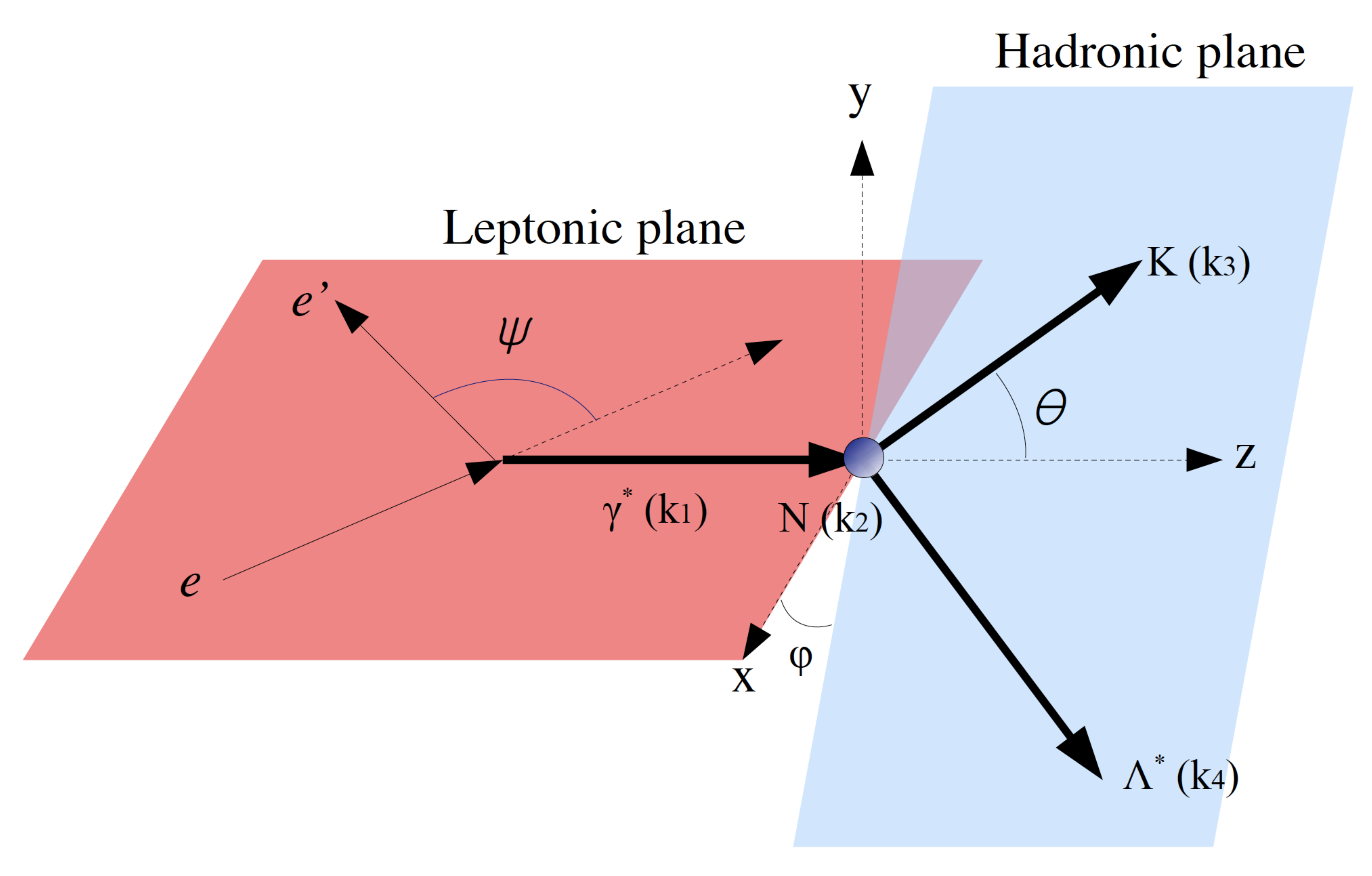}
\caption{(Color online) Definition of the leptonic and hadronic reaction planes for $eN\to e'K\Lambda^*$. Note that $\varphi$ indicates the azimuthal angle.}       
\label{FIG1}
\end{figure}

%FIGURE<<<
In this Section, we introduce the theoretical framework to compute the $\Lambda^*$ electroproduction off the nucleon target. We note that all the calculations are performed in the $K$-$\Lambda^*$ cm frame, where the four momenta of the particles, depicted in Figure~\ref{FIG0} for $\gamma^*N\to K\Lambda^*$ and Figure~\ref{FIG1} for $eN\to e'K\Lambda^*$, are defined as follows:
%EQUATION>>>
\begin{eqnarray}
\label{eq:MOMENTA}
\label{eq:APP1}
k_1=(E_1,0,0,k),\,\,
k_2=(E_2,0,0,-k),\,\,
k_3=(E_3,p\sin\theta,0,p\cos\theta),\,\,
k_4=(E_4,-p\sin\theta,0,-p\cos\theta),
\end{eqnarray}
%EQUAITON<<<
where $k_1$, $k_2$, $k_3$, and $k_4$ stand for the four momenta for the incident virtual photon, target nucleon, outgoing kaon, and recoil $\Lambda^*$, and the same for their masses $M_{1\sim4}$, i.e. $M_{1,2,3,4}=M_{\gamma,N,K,\Lambda^*}$. $\theta$ for the angle between the photon and the kaon in the cm frame. As understood, the reaction plane is defined by the $x$-$z$ plane, whereas $z$ direction is set to be parallel to the incident virtual photon three momentum. In the electroproduction, there are two  {\it independent} kinematic variables $W^2$ and $Q^2$, which are defined by $W^2=(k_1+k_2)^2=(k_3+k_4)^2$ and $Q^2=-k^2_1>0$, respectively. Hereafter, we employ a notation $\sqrt{Q^2}\equiv|\bm{q}|$ for convenience. Using these variables, the energies of the particles can be written by
%EQUATION>>>
\begin{equation}
\label{eq:APP3}
E_1=\frac{W^2-Q^2-M^2_2}{2W},\,\,\,\,
E_2=\frac{W^2+Q^2+M^2_2}{2W}\,\,\,\,
E_3=\frac{W^2-M^2_4+M^2_3}{2W},\,\,\,\,
E_4=\frac{W^2+M^2_4-M^2_3}{2W}.
\end{equation}
%EQUAITON<<<
Here, we write the absolute values for the three momenta for the initial- and final-state particles:
%EQUATION>>>
\begin{equation}
\label{eq:APP4}
k=\sqrt{E^2_1+Q^2},
\,\,\,\,\,
p=\sqrt{E^2_3-M^2_3}.
\end{equation}
%EQUAITON<<<
Now, we are in a position to define the frame-independent transverse-polarization parameter $\varepsilon$, which measures the strength of the transverse polarization in the virtual photon:
%EQUATION>>>
\begin{equation}
\label{eq:TL}
\varepsilon=\left[1+\frac{2k^2_1}{Q^2}\tan^2\left(\frac{\Psi}{2}\right)\right]^{-1},
\end{equation}
%EQUAITON<<<
where $\Psi$ denotes the polar angle for the electron for $eN\to e'K\Lambda^*$ as depicted in Figure~\ref{FIG1}. Using $\varepsilon$, the photon-polarization vectors, two transverse $(x,y)$ and one longitudinal $(z)$, are given as follows~\cite{Akerlof:1967zza}:
%EQUATION>>>
\begin{equation}
\label{eq:POLAR}
\epsilon_x=\left(0,0,\sqrt{1-\varepsilon},0\right),\,\,\,\,
\epsilon_y=\left(0,\sqrt{1+\varepsilon},0,0\right),\,\,\,\,
\epsilon_z=\frac{\sqrt{2\varepsilon}}{|\bm{q}|}\left(k,0,0,E_1\right).
\end{equation}
%EQUAITON<<<
As shown in Ref.~\cite{Akerlof:1967zza}, the scalar ($0$-th) components of the scattering amplitude can be omitted by virtue of the gauge invariance (WT identity). According to this, the photon-polarization vectors can be modified into three vectors by omitting the $0$-th component and multiplying a factor $k^2_1/E^2_1$, resulting in $\varepsilon\to\varepsilon_L\equiv(Q^2/E^2_1)\varepsilon$ (see Appendix). We, however, do not consider this treatment in the present work, keeping all the Lorentz components intact in the calculations, satisfying the WT identity explicitly. For more details for the electroproduction of pseudoscalar mesons, one can refer Refs.~\cite{Brown:1973wr,DESY 74/45,Amaldi:1979vh,Williams:1992tp,Chiang:2001as}. In the center-of-mass (cm) frame, the cross section is defined as
%EQUATION>>>
\begin{equation}
\label{eq:TCSD}
\frac{d^2\sigma}{d\Omega_{K}}=\frac{d^2\sigma_T}{d\Omega_{K}}
+\varepsilon\frac{d^2\sigma_L}{d\Omega_{K}}
+\varepsilon\frac{d^2\sigma_{TT}}{d\Omega_{K}}\cos2\varphi
+\sqrt{\varepsilon(1+\varepsilon)}\,\frac{d^2\sigma_{LT}}{d\Omega_{K}}\cos\varphi,
\end{equation}
%EQUAITON<<<
where $\Omega_{K}$ indicates the solid angle for the outgoing $K$ in the cm frame. $\sigma_{T,L}$ denote the contributions from the  transverse ($T$) and  longitudinal ($L$) photon polarizations, and $\sigma_{TT,LT}$ stand for the interferences between them~\cite{Brown:1973wr,DESY 74/45,Amaldi:1979vh,Williams:1992tp,Chiang:2001as,Mart:2010ch}. The azimuthal angle $\varphi$ defines the angle between the leptonic and hadronic planes as shown in Figure~\ref{FIG1}. For the unpolarized cross section that we are interested in the present work, we consider only the first and second terms in the right-hand-side of Eq.~(\ref{eq:TCSD}) by integrating over $\varphi$, resulting in that the interference terms disappear.
%FIGURE>>>
%FIGURE<<<

The effective Lagrangians for the EM and strong interaction vertices are defined by
%EQUATION>>>
\begin{eqnarray}
\label{eq:GROUND}
{\cal L}_{\gamma KK}&=&
ie_K\left[(\partial^{\mu}K^{\dagger})K-(\partial^{\mu}K)K^{\dagger}
\right]A_{\mu}+{\rm h.c.},
\nonumber\\
{\cal L}_{\gamma NN}&=&-\bar{N}\left[e_N\rlap{\,/}{A}-\frac{e\kappa_N}{4M_N}\sigma\cdot F\right]N+{\rm h.c.},
\cr
\mathcal{L}_{\gamma\Lambda^*\Lambda^*}&=&
-\bar{\Lambda}^{*\mu}
\left[\left(-F_{1}\rlap{/}{\epsilon}g_{\mu\nu}+F_K\rlap{/}{\epsilon}\frac{k_{1
\mu}k_{1
\nu}}{2M^{2}_{\Lambda^*}}\right)-\frac{\rlap{/}{k}_{1}\rlap{/}{\epsilon}}
{2M_{\Lambda^*}}\left(-F_{2}g_{\mu\nu}+F_4\frac{k_{1\mu}k_{1 \nu}}
{2M^{2}_{\Lambda^*}}\right)\right]
\Lambda^{*\nu}+\mathrm{h.c.},
\cr
\mathcal{L}_{\gamma KK^{*}}&=&
g_{\gamma KK^{*}}\epsilon_{\mu\nu\sigma\rho}(\partial^{\mu}A^{\nu})
(\partial^{\sigma}K)K^{*\rho}+\mathrm{h.c.},
\cr
\mathcal{L}_{\gamma
KN\Lambda^*}&=&-\frac{ie_Ng_{KN\Lambda^*}}{M_{\Lambda^*}}
\bar{\Lambda}^{*\mu}
A_{\mu}K\gamma_5N+{\rm h.c.},
\cr
\mathcal{L}_{KN\Lambda^*}&=&\frac{g_{KN\Lambda^*}}{M_{\Lambda^*}}
\bar{\Lambda}^{*\mu}\partial_{\mu}K\gamma_5N\,+{\rm h.c.},
\cr
\mathcal{L}_{K^{*}N\Lambda^*}&=&
-\frac{iG_{1}}{M_{V}}\bar{\Lambda}^{*\mu}\gamma^{\nu}G_{\mu\nu}N
-\frac{G_{2}}{M^{2}_{V}}\bar{\Lambda}^{*\mu}
G_{\mu\nu}\partial^{\nu}N
+\frac{G_{3}}{M^{2}_{V}}\bar{\Lambda}^{*\mu}\partial^{\nu}G_{\mu\nu}N+\mathrm{h.c.}.
\end{eqnarray}
%EQUATION<<<
Here, $e_{h}$ and $e$ denote the electric charge of the hadron $h$ and unit electric charge, respectively. The $A$, $K$, $K^{*}$, $N$, and $\Lambda^{*}$ indicate the fields for the photon, kaon, vector kaon, nucleon, and $\Lambda^{*}$ fields, respectively. As for the spin-$3/2$ fermion field, we employ of the Rarita-Schwinger (RS) vector-spinor field~\cite{Rarita:1941mf,Nath:1971wp}.

We also make use of the notation $\sigma\cdot F=\sigma_{\mu\nu}F^{\mu\nu}$, where $\sigma_{\mu\nu}=i(\gamma_{\mu}\gamma_{\nu}-\gamma_{\nu}\gamma_{\mu})/2$ and the EM field strength tensor $F^{\mu\nu}=\partial^{\mu}A^{\nu}-\partial^{\nu}A^{\mu}$. $\kappa_{N,\Lambda^{*}}$ denote the anomalous magnetic moments for the nucleon and $\Lambda^{*}$. Although the spin-$3/2$ $\Lambda^{*}$ has four different electromagnetic form factors $F_{1,2,3,4}$ as shown in Eq.~(\ref{eq:GROUND}) in general, we only take into account the dipole one ($F_{2}\equiv e_{Q}\kappa_{\Lambda^{*}}$) as a free parameter, since we do not have any theoretical and experimental information for it. Hence, we ignore the monopole ($F_{1}\equiv e_{\Lambda^{*}}=0$), quadrupole ($F_{3}$),  and octupole ($F_{4}$) ones, since their contributions are assumed to be negligible. As a trial, we will choose $\kappa_{\Lambda^{*}}\sim\kappa_n$ in the present work. Using the $\gamma KK^{*}$ interaction given in Eq.~(\ref{eq:GROUND}) and experimental data~\cite{Amsler:2008zzb}, one obtains that $g_{\gamma K^{*\pm}K^{\mp}}=0.254/\mathrm{GeV}$ and $g_{\gamma K^{*0}K^0}=0.358/\mathrm{GeV}$. $g_{KN\Lambda^{*}}$ can be computed with the experimental data for the full and partial decay widths: $\Gamma_{\Lambda^{*}}\approx15.6$ MeV and $\Gamma_{\Lambda^{*}\to\bar{K}N}/\Gamma_{\Lambda^{*}}\approx0.45$~\cite{Amsler:2008zzb}, resulting in that $g_{KN\Lambda^*}\approx11$. As for the $K^{*}N\Lambda^{*}$ interaction, there are three individual terms, and we defined a notation $G_{\mu\nu}=\partial_{\mu}K^{*}_{\nu}-\partial_{\nu}K^{*}_{\mu}$. Since we have only insufficient experimental and theoretical information to determine all the coupling strengths for the $G_{1,2,3}$, we set $G_{2}$ and $G_{3}$ to be zero for simplicity. We will take the value for $G_{1}\equiv g_{K^{*}N\Lambda^{*}}$ from the theoretical estimation using the coupled chiral unitary approach, resulting in $|g_{K^{*}N\Lambda^{*}}|\approx1.5$~\cite{Hyodo:2006uw}. 

In the present work, we are interested in three nucleon resonances, i.e. $S_{11}(2090,J^P=1/2^-)$, $D_{13}(2080,J^P=3/2^-)$, and $D_{15}(2200,J^P=5/2^-)$~\cite{Nakamura:2010zzi}. Note that the contribution from $D_{13}$ was suggested to give a considerable contribution to the $\Lambda^*$ photoproduction~\cite{Kohri:2009xe,Xie:2010yk,He:2012ud}. The effective strong and EM Lagrangians for the $s$-channel diagram for the nucleon resonances read~\cite{Kim:2012pz}: 
%EQUATION>>>
\begin{eqnarray}
\label{eq:LAGR}
\mathcal{L}_{\gamma  NR_1}  &=& 
e \bar{N}\left[\frac{h_{11}}{2M_N}\gamma_5\sigma_{\mu\nu}
\partial^\nu -h_{31}\left(\frac{\gamma_\mu\partial^2}{M_N+M_{R_1}}+i\partial_\mu\right)
\gamma_5  \right]A^\mu R_1 + \mathrm{h.c.},                       
\cr
\mathcal{L}_{\gamma  NR_3} &=& -ie \left[ \frac{h_{13}}{2M_N}
\bar{N} \gamma^\nu F_{\mu\nu}-\frac{ih_{23}}{(2M_N)^2} (\partial^\nu \bar{N})F_{\mu\nu}
-\frac{ih_{33}}{(2M_N)^2} \bar{N}\partial^\nu(F_{\mu\nu}) \right] 
 R_3^{\mu}  + \mathrm{h.c.},                       
\cr
\mathcal{L}_{\gamma  NR_5}  &=& 
e\left[ \frac{h_{15}}{(2M_N)^2} \bar{N}\gamma_5\gamma^\nu\partial_\alpha F_{\mu\nu}
-\frac{ih_{25}}{(2M_N)^3}(\partial^\nu\bar{N})\gamma_5\partial_\alpha F_{\mu\nu}-\frac{ih_{35}}{(2M_N)^3}\bar{N}\gamma_5(\partial^\nu\partial_\alpha F_{\mu\nu}) \right] 
R_5^{\mu\alpha} + \mathrm{h.c.},
\cr
\mathcal{L}_{KR_1\Lambda^*}&=&\frac{g_{11}}{M_K}\bar{\Lambda}^*_\mu
(\partial^\mu K)R_1+\mathrm{h.c.},
\cr
\mathcal{L}_{KR_3\Lambda^*}&=&\frac{g_{13}}{M_K}\bar{\Lambda}^*_\mu\gamma_5(\rlap{/}{\partial}K)R_3^{\mu}+\frac{ig_{23}}{M^2_K}\bar{\Lambda}^*_\mu\gamma_5(\partial^\mu\partial_\nu K)R_3^{\nu}+\mathrm{h.c.},
\cr
\mathcal{L}_{KR_5\Lambda^*}&=&\frac{ig_{15}}{M^2_K}\bar{\Lambda}^*_\mu
(\partial_\sigma\partial_\nu K)\gamma^\sigma R^{\mu\nu}_5
-\frac{g_{25}}{M^3_K}\bar{\Lambda}^*_\mu
(\partial^\mu\partial_\nu\partial_\sigma K)R^{\nu\sigma}_5+\mathrm{h.c.},
\end{eqnarray}
%EQUAITON<<<
where $R_{1,3,5}$  stand for the resonance fields of  $S_{11}(2090)$, $D_{13}(2080)$, and $D_{15}(2200)$, respectively. As already mentioned, we employed the RS vector-spinor formalism here for them. Note that, in general, the helicity amplitude for $\gamma^*N\to R$ is given by the transverse ($A_{1/2}$ and $A_{3/2}$) and longitudinal ($A_{0}$) ones, and they can defined by~\cite{Drechsel:1998hk} 
%EQUATION>>>
\begin{equation}
\label{eq:TLHA}
A^R_{1/2}=\sqrt{\frac{\pi\alpha_\mathrm{EM}}{k_W}}\langle R,\frac{1}{2},|J_x+i J_y|N,-\frac{1}{2}\rangle\,\,\,\,
A^R_{3/2}=\sqrt{\frac{\pi\alpha_\mathrm{EM}}{k_W}}\langle R,\frac{3}{2},|J_x+i J_y|N,\frac{1}{2}\rangle\,\,\,\,
A^R_{0}=\sqrt{\frac{2\pi\alpha_\mathrm{EM}}{k_W}}
\langle R,\frac{1}{2},|J_0|N,\frac{1}{2}\rangle,
\end{equation}
%EQUAITON<<<
where $\alpha_\mathrm{EM}$ stands for the EM fine-structure constant, and we have ignored the overall phase factor for simplicity. $J_\mu$ and $k_W$ stands for the EM current and ${(W^2-M^2_N)}/{(2W)}$. In comparison to the transverse helicity amplitudes, the longitudinal helicity amplitude has not been well explored experimentally and theoretically~\cite{Santopinto:2012nq,DeSanctis:1998tu}. Moreover, there is only insufficient information for $A^{R}_{0}$, and to compute them employing a different theoretical model is beyond our scope of the present work. Hence, taking into account the present situation, we simply set $h_{31,33,35}$ to be zero in Eq.~(\ref{eq:LAGR}), ignoring the $A^R_{0}$ contribution for brevity, throughout the present work as in Ref.~\cite{Drechsel:1998hk}. Then, $A^R_{1/2}$ and $A^R_{3/2}$ can be straightforwardly evaluated using the Lagrangians defined in Eq.~(\ref{eq:LAGR}) and Eq.~(\ref{eq:TLHA}), then we obtain the following expressions for the transverse helicity amplitudes to determined the transition coupling strengths~\cite{Oh:2007jd}:
%EQUATION>>>
\begin{eqnarray}
\label{eq:HEL}
A^{R_1}_{1/2}&=&-\frac{eh_{11}}{2M_N}\sqrt{\frac{k_\gamma M_{R_1}}{M_N}},
\cr
A^{R_3}_{1/2}&=&\frac{e\sqrt{6}}{12}\sqrt{\frac{k_\gamma}{M_NM_{R_3}}}
\left[h_{13}+\frac{h_{23}}{4M^2_N}M_{R_3}(M_{R_3}+M_N) \right],
\cr
A^{R_3}_{3/2}&=&\frac{e\sqrt{2}}{4M_N}\sqrt{\frac{k_\gamma M_{R_3}}{M_N}}
\left[h_{13}+\frac{h_{23}}{4M_N}(M_{R_3}+M_N) \right],
\cr
A^{R_5}_{1/2}&=&-\frac{e}{4\sqrt{10}}\frac{k_\gamma}{M_N}
\sqrt{\frac{k_\gamma}{M_NM_{R_5}}}
\left[h_{15}+\frac{h_{25}}{4M^2_N}M_{R_5}(M_{R_5}-M_N) \right],
\cr
A^{R_5}_{3/2}&=&-\frac{e}{4\sqrt{5}}\frac{k_\gamma}{M^2_N}\sqrt{\frac{k_\gamma M_{R_5}}{M_N}}
\left[h_{15}-\frac{h_{25}}{4M_N}(M_{R_5}-M_N) \right].
\end{eqnarray}
%EQUAITON<<<
Using the experimental~\cite{Nakamura:2010zzi} and theoretical ~\cite{Capstick:1992uc} informations for the helicity amplitudes for $D_{13}$ and $(S_{11},D_{15})$, respectively, as given in Table~\ref{TABLE1}, we have the following values for the proton $(p)$ and neutron $(n)$ resonances, ignoring the errors:
%EQUATION>>>
\begin{eqnarray}
\label{eq:HR}
h^{(p,n)}_{11}&=&(-0.055,+0.018),\,\,\,\,h^{(p,n)}_{31}=0,
\cr
h^{(p,n)}_{13}&=&(+0.608,-0.770),\,\,\,\,h^{(p,n)}_{23}=(-0.620,+0.531),
\,\,\,\,h^{(p,n)}_{33}=0,
\cr
h^{(p,n)}_{15}&=&(+0.123,-0.842),\,\,\,\,h^{(p,n)}_{25}=(+0.011,-0.872),
\,\,\,\,h^{(p,n)}_{35}=0.
\end{eqnarray}
%EQUAITON<<<

As for the strong couplings for the $KR\Lambda^*$ vertex, we have not had experimental as well as theoretical information yet. Instead, there were the SU(6) quark-model calculations for them~\cite{Capstick:1998uh}. The partial decay amplitude for $R\to K\Lambda^*$ is related to the $G(\ell)$ through the following equation~\cite{Capstick:1998uh}: 
%EQUATION>>>
\begin{equation}
\label{eq:GAMMASTR}
\Gamma_{R\to K\Lambda^*}=\sum_\ell|G(\ell)|^2.
\end{equation}
%EQUAITON<<<
Since we are interested in the relatively low-energy region, i.e. $|\bm{k}_K|\ll M_K$, we rather safely ignore the second terms by setting $g_{23,25}=0$ in Eq.~(\ref{eq:LAGR})~\cite{Kim:2011rm}. Employing the following equation for determining the strong couplings~\cite{Toki:2007ab,Oh:2007jd}
%EQUATION>>>
\begin{eqnarray}
\label{eq:GGGG}
\Gamma_{R_1\to K\Lambda^*}&\approx&\frac{g^2_{11}\,|\bm{q}|^3M_{R_1}\,
(E_{\Lambda^*}+M_{\Lambda^*})}{3\pi M^2_{\Lambda^*}M^2_K},
\cr
\Gamma_{R_3\to K\Lambda^*}&\approx&\frac{g^2_{13}\,|\bm{q}|\,
(E_{\Lambda^*}-M_{\Lambda^*})}{18\pi M_{R_{3}}M^2_K}
\left[\frac{M_{R_3}+M_{\Lambda^*}}{M_{\Lambda^*}} \right]^2
\left[E^2_{\Lambda^*}-E_{\Lambda^*}M_{\Lambda^*}+\frac{5}{2}M^2_{\Lambda^*} \right],
\cr
\Gamma_{R_5\to K\Lambda^*}&\approx&\frac{2g^2_{15}\,|\bm{q}|^3\,
(E_{\Lambda^*}+M_{\Lambda^*})}{45\pi M_{R_5}M^4_K}
\left[\frac{M_{R_3}-M_{\Lambda^*}}{M_{\Lambda^*}} \right]^2
\left[E^2_{\Lambda^*}+E_{\Lambda^*}M_{\Lambda^*}+\frac{7}{4}M^2_{\Lambda^*} \right],
\end{eqnarray}
%EQUAITON<<<
where $\Gamma_{R\to K\Lambda^*}$ indicates the strong partial decay width for the resonance $R$. The absolute value for the thee momentum for the decaying particle can be computed by the K\"allen function
%EQUATION>>>
\begin{equation}
\label{eq:KAL}
|\bm{q}|=\frac{\sqrt{[M^2_{R}-(M_{\Lambda^*}+M_K)^2][M^2_{R}-(M_{\Lambda^*}-M_K )^2 ]}}{2M_{R}},
\end{equation}
%EQUAITON<<<
and using the values given in Ref.~\cite{Capstick:1998uh}, we have the following center values for the strong coupling constants, considering the assumption:
%EQUATION>>>
\begin{equation}
\label{eq:STONGC}
|g_{11}|=1.53,\,\,\,\,|g_{13}|=1.25,\,\,\,\,|g_{15}|=0.40,\,\,\,\,g_{23,25}=0.
\end{equation}
%EQUAITON<<<
Here, we note that the nucleon resonances in the present work are identified with those in the SU(6) quark model~\cite{Capstick:1998uh} as follows:
%EQUATION>>>
\begin{equation}
\label{eq:SM}
S_{11}(2090)\leftrightarrow[N\,1/2^-]_3(1945),\,\,\,\,
D_{13}(2080)\leftrightarrow[N\,3/2^-]_3(1960),\,\,\,\,
D_{15}(2200)\leftrightarrow[N\,5/2^-]_3(2095).
\end{equation}
%EQUAITON<<<
%TABLE>>>
\begin{table}[b]
\begin{tabular}{c|c|c|c|c|c|c}
$R$&QM~\cite{Capstick:1998uh}&$\Gamma_R$ [MeV]&$A^{R\to\gamma(p,n)}_{1/2}$ [$\frac{1}{\sqrt{\mathrm{GeV}}}$]&$A^{R\to\gamma(p,n)}_{3/2}$ [$\frac{1}{\sqrt{\mathrm{GeV}}}$]&$\Gamma_{R\to K\Lambda^*}$ [MeV]&$G(1)$ [$\sqrt{\mathrm{MeV}}$]\\
\hline
$S_{11}(2090)$&$[N\frac{1}{2}^-]_3(1945)$&$250$&$(+0.012,-0.004)$&$-$&$40.96$&$+6.4^{+5.7}_{-6.4}$\\
$D_{13}(2080)$&$[N\frac{3}{2}^-]_3(1960)$&$250$&$(-0.020,+0.007)$&$(+0.017,-0.053)$&$6.76$&$-2.6^{+2.6}_{-2.8}$\\
$D_{15}(2200)$&$[N\frac{5}{2}^-]_3(2095)$&$250$&$(-0.002,+0.022)$&$(-0.006,+0.029)$&$5.76$&$-2.4^{+2.4}_{-2.0}$\\
\end{tabular}
\caption{Input parameters for $S_{11}(2090)$, $D_{13}(2080)$, and $D_{15}(2200)$ taken from the experimental and theoretical estimation~\cite{Capstick:1992uc,Nakamura:2010zzi,Capstick:1998uh}}
\label{TABLE1}
\end{table}
%TABLE>>>

Using all the effective Lagrangians given in Eqs.~(\ref{eq:GROUND}) and (\ref{eq:LAGR}), one can straightforwardly evaluate the following invariant amplitudes, corresponding to the Feynman diagrams in Figure~\ref{FIG0}:
%EQUATION>>>
\begin{eqnarray}
\label{eq:AMP}
i\mathcal{M}^N_{s}&=&-\frac{g_{KN\Lambda^*}}{M_{\Lambda^*}}
\bar{u}^{\mu}_2k_{3\mu}{\gamma}_{5}
\left[\frac{e_{N}[\rlap{/}{k}_{1}F_N(s)+(\rlap{/}{k}_{2}+M_{N})F_{c}(s,t)]}
{s-M^{2}_{N}}\rlap{/}{\epsilon}
-\frac{e_{Q}\kappa_{p}}{2M_{N}}
\frac{(\rlap{/}{k}_{1}+\rlap{/}{k}_{2}+M_{N})F_N(s)}
{s-M^{2}_{N}}\rlap{/}{\epsilon}\rlap{/}{k}_{1}
\right]u_1,
\cr
i\mathcal{M}^{\Lambda^*}_{u}&=&-\frac{e_{Q}g_{KN\Lambda^*}\kappa_{\Lambda^*}}
{2M_{\Lambda^*}M_{\Lambda}}
\bar{u}^{\mu}_{2}
\left(\rlap{/}{k}_{1}\rlap{/}{\epsilon} \right)
\left[\frac{(\rlap{/}{k}_{4}-\rlap{/}{k}_{1}+M_{\Lambda^{*}})}
{u-M^{2}_{\Lambda^{*}}}
 \right]k_{3\mu}\gamma_{5}u_{1} F_{\Lambda^*}(u),
\cr
i\mathcal{M}^{K}_{t}&=&\frac{e_{K}g_{KN\Lambda^*}}{M_{\Lambda^*}}
\bar{u}^{\mu}_2
\left[\frac{(k_{1\mu}-k_{3\mu})[2(\epsilon\cdot k_3)F_c(s,t)-(\epsilon\cdot k_1)F_K(t)]}
{t-M^{2}_{K}} \right]
\gamma_{5}u_1 ,
\cr
i\mathcal{M}^{K^{*}}_{t}&=&
-\frac{ig_{\gamma{K}K^*}g_{K^{*}NB}}{M_{K^{*}}}
\bar{u}^{\mu}_{2}\gamma_{\nu}
\left[\frac{(k^{\mu}_{1}-k^{\mu}_{3})g^{\nu\sigma}-
(k^{\nu}_{1}-k^{\nu}_{3})g^{\mu\sigma}}{t-M^{2}_{K^*}}\right]
(\epsilon_{\rho\eta\xi\sigma}k^{\rho}_{1}
\epsilon^{\eta}k^{\xi}_{3})u_1 F_{K^*}(t),
\cr
i\mathcal{M}_\mathrm{gauge}&=&
-\frac{g_{KN\Lambda^*}}{M_{\Lambda^*}}
\bar{u}^{\mu}_2{\gamma}_{5}(\epsilon\cdot k_1)
\left[\frac{e_Nk_{3\mu}[F_c(s,t)-F_N(s)]}{s-M^2_N}
+\frac{e_K(k_{1\mu}-k_{3\mu})[F_c(s,t)-F_K(t)]}{t-M^2_K}\right]u_1,
\cr
i\mathcal{M}_{\mathrm{contact}}
&=&\frac{e_{K}g_{KN\Lambda^*}}{M_{\Lambda^*}}
\bar{u}^{\mu}_2\epsilon_{\mu}{\gamma}_{5}u_1 F_c(s,t),
\cr
i\mathcal{M}^{R_1}_{s}&=&\frac{e|g_{11}|h_{11}e^{i\phi_1}}{4M_KM_N}\bar{u}_2^\mu\frac{k^\mu_3(\rlap{/}{k}_1+\rlap{/}{k}_2+M_N)}{s-M^2_{R_1}+i\Gamma_{R_1}M_{R_1}}
\left(\rlap{/}{\epsilon}\rlap{/}{k}_1-\rlap{/}{k}_1\rlap{/}{\epsilon} \right)\gamma_5u_1F_R(s).
\cr
i\mathcal{M}^{R_3}_{s}&=&\frac{e|g_{13}|e^{i\phi_3}}{2M_KM_N}\bar{u}_2^\mu\frac{\gamma_5\rlap{/}{k}_3(\rlap{/}{k}_1+\rlap{/}{k}_2+M_N)\mathcal{G}^{{R_3}}_{\mu\nu}}{s-M^2_{R_3}+i\Gamma_{R_3}M_{R_3}}
\left[h_1(k_1^\nu\rlap{/}{\epsilon}-\rlap{/}{k}_1\epsilon^{\nu})-\frac{h_2}{2M_N}
[k_1^\nu(\epsilon\cdot k_2)-\epsilon^\nu(k_1\cdot k_2)] \right]u_1F_R(s).
\cr
i\mathcal{M}^{R_5}_{s}&=&\frac{e|g_{15}|e^{i\phi_5}}{4M^2_KM^2_N}\bar{u}_2^\mu\frac{\rlap{/}{k}_3(\rlap{/}{k}_1+\rlap{/}{k}_2+M_N)k^{\nu}_3\mathcal{G}^{R_5}_{\mu\nu\sigma\rho}k^{\rho}_1}{s-M^2_{R_5}+i\Gamma_{R_5}M_{R_5}}
\left[h_1(k_1^\sigma\rlap{/}{\epsilon}-\rlap{/}{k}_1\epsilon^{\sigma})-\frac{h_2}{2M_N}
[k_1^\sigma(\epsilon\cdot k_2)-\epsilon^\sigma(k_1\cdot k_2)] \right]\gamma_5u_1F_R(s),
\nonumber\\
\label{amplitudes}
\end{eqnarray}
%EQUAITON<<<
where $F_h$ and $\Gamma_R$ stand for the EM-strong form factor and full decay width for the resonance. Note that we have employed the phase factor for the resonances $e^{i\phi_{1,3,5}}$, since we can not determine the phase for the resonances within the present model. Hence, these phase angle $\phi$ will be determined to reproduce the experimental data. The spin-$(3/2,5/2)$ projection operators can be written as follows:
%EQUATION>>>
\begin{eqnarray}
\label{eq:PROJ}
\mathcal{G}^{R_3}_{\mu\nu}&=&g_{\mu\nu}-\frac{1}{3}\gamma_\mu\gamma_\nu-\frac{1}{3M_R}
(\gamma_\mu q_{\nu}-\gamma_\nu q_{\mu})-\frac{2}{3M^2_R}q_{\mu}q_{\nu},
\cr
\mathcal{G}^{R_5}_{\mu\nu\sigma\rho}&=&
\frac{1}{2}( \bar g_{\mu\sigma} \bar g_{\nu\rho} 
+\bar g_{\mu\rho} \bar g_{\nu\sigma})
-\frac{1}{5} \bar g_{\mu\nu} \bar g_{\sigma\rho} 
-\frac{1}{10}( \bar \gamma_\mu \bar \gamma_\sigma \bar
g_{\nu\rho}  
+\bar \gamma_\mu \bar \gamma_\rho \bar g_{\nu\sigma}
+\bar \gamma_\nu \bar \gamma_\sigma \bar g_{\mu\rho} 
+\bar \gamma_\nu \bar \gamma_\rho \bar g_{\mu\sigma} ) ,
\end{eqnarray}
%EQUAITON<<<
where we have used the notations for simplicity: 
%EQUATION>>>
\begin{equation}
\label{eq:NONONO}
\bar{g}_{\mu\nu}\equiv g_{\mu\nu}-\frac{q_\mu q_\nu}{M^2_{R_5}},\,\,\,\,\,
\bar{\gamma}_\mu\equiv\gamma_\mu-\frac{q_\mu}{M^2_{R_5}}\rlap{/}{q},\,\,\,\,\,
q=k_1+k_2.
\end{equation}
%EQUAITON<<<
All the discussed input parameters for the resonances for the numerical calculations are listed in Table~\ref{TABLE1}.

The sum of all the invariant amplitudes satisfy the gauge invariance (WT identity), $k_1\cdot\mathcal{M}_\mathrm{total}=0$, with the phenomenological form factors $F_{h}$, following the gauge-conserving form-factor prescription suggested by Refs.~\cite{Haberzettl:1998eq,Davidson:2001rk,Haberzettl:2006bn}. Note that we have employed an additional term $\mathcal{M}_\mathrm{gauge}$ to conserve the gauge invariance of the scattering amplitude. This choice of $\mathcal{M}_\mathrm{gauge}$ is slightly different from the usual pseudoscalar meson electroproduction with the ground state baryons~\cite{Mart:2010ch}, in which a term proportional to $(k_1\cdot\epsilon)/Q^2$ is taken into account to save the gauge invariance. We, however, consider that the electroproduction scattering amplitude should be smoothly interpolated as $Q^2\to0$ to that for the photoproduction, which is given in our previous work~\cite{Nam:2005uq,Nam:2009cv,Nam:2010au} and provided good agreement with experimental data~\cite{Barber:1980zv,Muramatsu:2009zp}. In this sense, we introduce the additional term, $\mathcal{M}_\mathrm{gauge}$ for the gauge invariance as in Eq.~(\ref{eq:AMP}) from a phenomenological point of view. It is worth mentioning that this term does not make any effect on the physical observable, due to that $k_1\cdot\epsilon=0$ for any photon polarization. The form factors $F_h$ are parameterized by:
%EQUATION>>>
\begin{equation}
\label{eq:FORM}
F_{N,R}(s)=\frac{\Lambda^{4}_h}{\Lambda^{4}_h+(s-M^{2}_{N,R})^{2}},\,\,\,\,
F_{\Lambda^*}(u)=\frac{\Lambda^{4}_h}{\Lambda^{4}_h+(u-M^{2}_{\Lambda^*})^{2}},
\,\,\,\,
F_K(t)=\left[\frac{\Lambda^2_h-M^2_K}{\Lambda^2_h-t} \right]^2,
\end{equation}
%EQUAITON<<<
where $x$ denote the Mandelstam variables. The {\it common} form factor to conserve the on-shell condition, i.e. a form factor becomes unity at zero photon virtuality, is assigned by~\cite{Davidson:2001rk,Nam:2005uq,Nam:2009cv,Nam:2010au}
%EQUATION>>>
\begin{equation}
\label{eq:fc}
F_{c}(s,t)=F_N(s)+F_K(t)-F_N(s)F_K(t).
\end{equation}
%EQUAITON<<<

\begin{figure}[t]
\includegraphics[width=8.5cm]{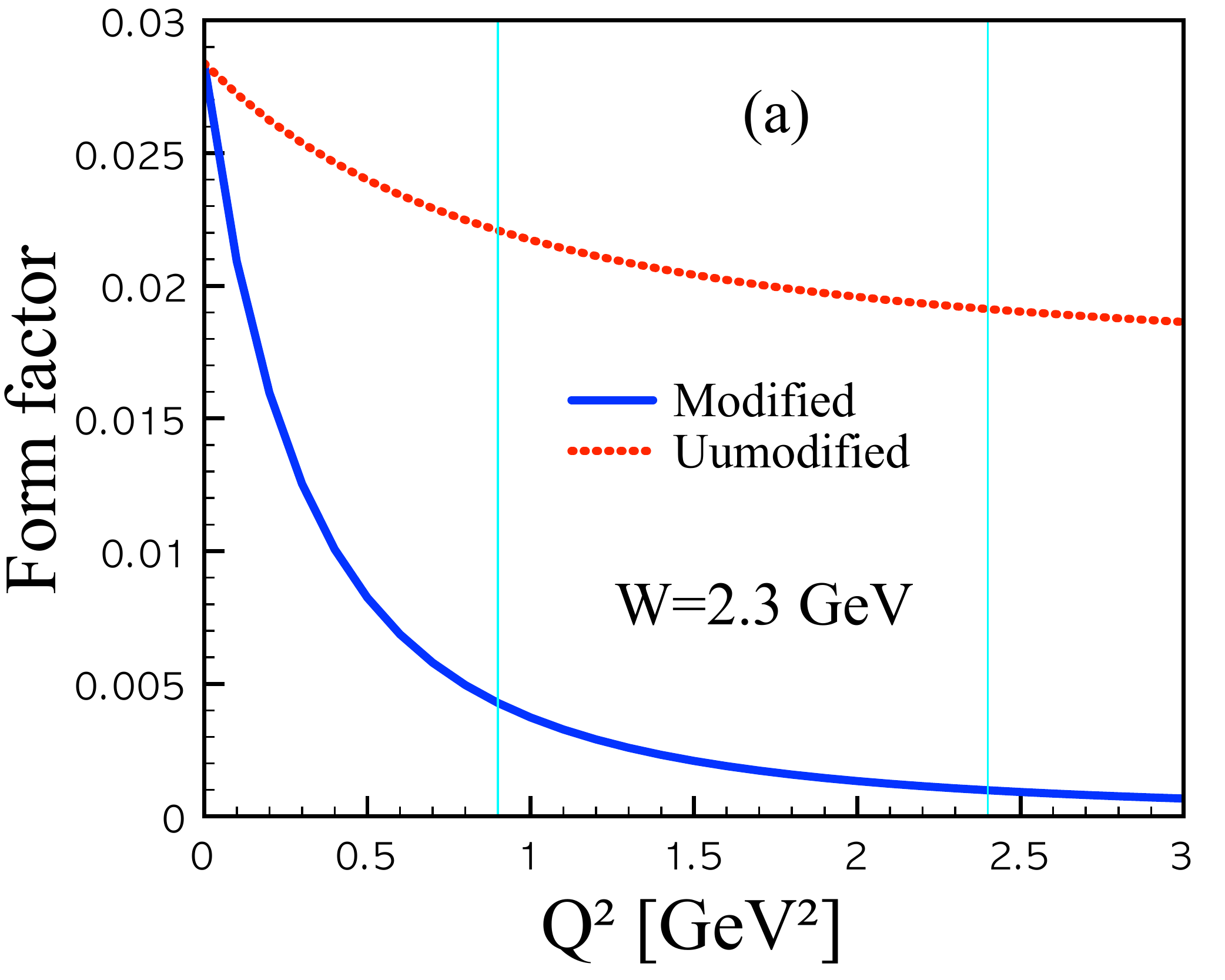}
\caption{(a) Form factors, $F_c(s,t)$ (unmodified; solid) and $F_c(s,t)F_\mathrm{DP}(Q^2)$ (modified; dot), as functions of $Q^2$ at $W=2.3$ GeV. The vertical lines indicate the experimentally possible region for the photon virtuality~\cite{Barrow:2001ds}: $Q^2=(0.9\sim2.4)\,\mathrm{GeV}^2$.}       
\label{FIG01}
\end{figure}

Here, we provide some discussions on the present form factor scheme. Note that the form factors in the above prescription was employed for  the $\Lambda(1520)$ photoproduction~\cite{Nam:2005uq,Nam:2009cv,Nam:2010au}. Hence, it is possible that the above form factor scheme is improper in describing the $Q^2$ dependence. Now, we consider an prescription to realize the proper  $Q^2$ dependence for the above form factors. The proton EM (Dirac) and charged kaon form EM factors are parameterized frequently with the phenomenologically well-established dipole- and monopole-type ones: 
%EQUATION>>>
\begin{equation}
\label{eq:DPFF}
F_\mathrm{EM}^p(Q^2)=\frac{1}{[1+Q^{2}/\Lambda^{2}_{N,\mathrm{EM}}]^2},
\,\,\,\,
F^{K^+}_{\mathrm{EM}}(Q^2)=\frac{1}{1+Q^{2}/\Lambda^{2}_{K,\mathrm{EM}}}.
\end{equation}
%EQUAITON<<<
Considering that the these form factor become unity at $Q^2\to0$, one can expect the form factor scheme for the photoproduction is modified for the electroproduction with this $Q^2$ dependence as
%EQUATION>>>
\begin{equation}
\label{eq:MODFF}
F_h(x)\to F_h(x)F^h_\mathrm{EM}(Q^2)\equiv F^\mathrm{mod}_h(x,Q^2).
\end{equation}
%EQUAITON<<<
In other words, at $Q^2=0$, the photoproduction amplitude is recovered from that for the electroproduction. By doing this, one can approximately describe the $Q^2$ dependence for the form factors. In the panel (a) of Figure~\ref{FIG01}, we depict $F^\mathrm{mod}_h(s,t,Q^2)$ (modified; solid) and $F_c(s,t)$ (unmodified; dot), as functions of $Q^2$ at $W=2.3$ GeV, which is the center value experimentally possible at the CLAS experiment~\cite{Barrow:2001ds}, since the effect of $F_c(s,t)$ dominates the process. Here, we choose $(\Lambda,\Lambda_\mathrm{DP})=(600,840)$ MeV as a trial. The vertical lines indicate the experimentally possible region for the photon virtuality: $Q^2=(0.9\sim2.4)\,\mathrm{GeV}^2$~\cite{Barrow:2001ds}. We observe that the quite different slope profiles for the two form factors even for $Q^2=(0.9\sim2.4)\,\mathrm{GeV}^2$. It also turns out that there appears a huge difference in their strengths. Hence, we can conclude that the modification of the form factors are inavoidable to describe the $Q^2$ dependence appropriately by using $F^\mathrm{mod}_h(x,Q^2)$ instead of $F_h(x)$. The cutoff mass $\Lambda$ will be determined to reproduce the data in the next Section. As for the resonances $p^*$ and $n^*$, we also modify $F_R\to F^\mathrm{mod}_R$ in Eq.~(\ref{eq:AMP}), although the $Q^2$ dependence for the resonance can be different from the nucleon, considering the lack of theoretical and experimental information.
%FIGURE>>>
%FIGURE<<<

%TABLE>>>
\begin{table}[b]
\begin{tabular}{c|c|c|c|c|c|c|c}
$\hspace{0.3cm}|g_{11}|\hspace{0.3cm}$&
$\hspace{0.3cm}|g_{13}|\hspace{0.3cm}$&
$\hspace{0.3cm}|g_{15}|\hspace{0.3cm}$
&$h^{p,n}_{11}$
&$h^{p,n}_{13}$ 
&$h^{p,n}_{15}$ 
&$h^{p,n}_{23}$ 
&$h^{p,n}_{25}$ \\
\hline
$1.53$&$1.25$&$0.40$&$(-0.055,+0.018)$&$(+0.608,-0.770)$&$(+0.123,-0.842)$&$(-0.620,+0.531)$&$(+0.011,-0.872)$\\
\end{tabular}
\caption{Electromagnetic and strong coupling constants for the resonances in the present work.}
\label{TABLE2}
\end{table}
%TABLE>>>

It is worth mentioning that, in Refs.~\cite{Pascalutsa:1998pw,Pascalutsa:1999zz,Pascalutsa:2000kd}, the authors discussed a condition that the free and interacting Lagrangians for the spin-$3/2$ fermion in terms of the RS formalism can be constructed {\it consistently} with the same numbers of the constraints, which cancel the redundant unphysical spin-$1/2$ components in the formalism. The condition can be achieved by constructing the interacting Lagrangian in a gauge-invariant (GI) way:
%EQUATION>>>
\begin{equation}
\label{eq:PASCAL}
\mathcal{L}^\mathrm{GI}_{KN\Lambda^*}=\frac{g^\mathrm{GI}_{KN\Lambda^*}}
{M^2_{\Lambda^*}}\epsilon^{\mu\nu\sigma\rho}(\partial_\mu\Lambda^*_\nu)
\gamma_\sigma
\partial_\rho K N\,+{\rm h.c.},
\end{equation}
%EQUAITON<<<
which survives in the tree-level $s$-channel diagram. In what follows, we assign the cases employing the Lagrangian in Eq.~(\ref{eq:PASCAL}) as a consistent treatment. Note that the form of the interacting Lagrangian $\mathcal{L}_{KN\Lambda^*}$ in Eq.~(\ref{eq:GROUND}) does not satisfies this condition so that it is called {\it inconsistent}.  In Ref.~\cite{Shklyar:2004dy}, the $\pi N$ elastic scattering with the higher-spin resonances was investigated by comparing the consistent and inconsistent treatments, resulting in that two different prescriptions present qualitatively equivalent results from a phenomenological point of view. If we use the consistent interacting Lagrangian in Eq.~(\ref{eq:PASCAL}), the invariant amplitudes for the $\Lambda^*$ EM production become those with the following change, satisfying the WT identity:
%EQUATION>>>
\begin{equation}
\label{eq:CHAN}
\frac{g_{KN\Lambda^*}}{M_{\Lambda^*}}v^\mu\gamma_5
\to \frac{ig^\mathrm{GI}_{KN\Lambda^*}}{M^2_{\Lambda^*}}k_{4\nu}\gamma_\sigma v_{\rho}\epsilon^{\mu\nu\sigma\rho},
\end{equation}
%EQUAITON<<<
where the four vector $v$ denotes the four momenta of the particles or photon polarization vector. As for the contact-term contribution, which is the main source for the production rate in the inconsistent treatment,  one has the following amplitudes for each case:
%EQUATION>>>
\begin{equation}
\label{eq:EACH}
\mathcal{M}_{\mathrm{contact}}
=\frac{e_{K}g_{KN\Lambda^*}}{M_{\Lambda^*}}
\bar{u}^{\mu}_2\epsilon_{\mu}{\gamma}_{5}u_1,\,\,\,\,
\mathcal{M}^\mathrm{GI}_{\mathrm{contact}}
=\frac{ie_{K}g^\mathrm{GI}_{KN\Lambda^*}}{M^2_{\Lambda^*}}
\bar{u}^{\mu}_2k^\nu_4\gamma^\sigma \epsilon^{\rho}\epsilon_{\mu\nu\sigma\rho}u_1.
\end{equation}
%EQUAITON<<<
From these inconsistent ($\mathcal{M}_{\mathrm{contact}}$) and consistent ($\mathcal{M}^\mathrm{GI}_{\mathrm{contact}}$) amplitudes for the $\Lambda^*$ photoproduction, one is led to 
%EQUATION>>>
\begin{eqnarray}
\label{eq:APP}
|\mathcal{M}_{\mathrm{contact}}|^2&=&\frac{4e^2_{K}g^2_{KN\Lambda^*}}{M^2_{\Lambda^*}}(k_N\cdot k_{\Lambda^*}-M_NM_{\Lambda^*})(\epsilon_{\Lambda^*}\cdot\epsilon_\gamma)^2,
\cr
|\mathcal{M}^\mathrm{GI}_{\mathrm{contact}}|^2&\approx&
\frac{4e^2_{K}(g^\mathrm{GI}_{KN\Lambda^*})^2}{M^2_{\Lambda^*}}(k_N\cdot k_{\Lambda^*}-M_NM_{\Lambda^*})
\left[(\epsilon_{\Lambda^*}\cdot\epsilon_\gamma)^2
%+\frac{(k_{\Lambda^*}\cdot\epsilon_\gamma)^2}{M^2_{\Lambda^*}}
-1\right],
\end{eqnarray}
%EQUAITON<<<
where $\epsilon_{\Lambda^*}$ denotes the spin-$1$ components of the RS field for $\Lambda^*$ and $k_{N,\Lambda^*}\equiv k_{2,4}$. As understood by seeing Eq.~(\ref{eq:APP}), in the consistent treatment, we have the additional  term in comparison to the inconsistent one. Although we do not perform quantitative calculations using the consistent treatment in the present work, we consider that the contact-term dominance~\cite{Nam:2005uq} can be modified by the possible cancelation appearing in the term $(\epsilon_{\Lambda^*}\cdot\epsilon_\gamma)^2-1$ in the consistent treatment. This observation can be interpreted as the cancelation of unphysical spin components of the inconsistent treatment. We want to leave this interesting issue with the consistent treatment as a future work.

%--------------------------------------------------
\section{Numerical Results}
%--------------------------------------------------
In this Section, we will provide and discuss various theoretical results for $\gamma^*N\to K\Lambda^*$. As for the experimental data to be compared with the numerical results, we will closely explore those from the experiment done by Barrow {\it et al.} of the CLAS collaboration~\cite{Barrow:2001ds}. In this experiment, the kinematical ranges for $W$ and $Q^2$ are $(1.95\sim2.65)$ GeV and $(0.9\sim2.4)\,\mathrm{GeV}^2$, respectively. Moreover, the value of $\varepsilon$, defined in Eq.~(\ref{eq:TL}),  ranges from $0.3$ to $0.7$, depending on $W^2$ and $Q^2$. Considering these experimental conditions, for the numerical calculations, we will make use of the value $\varepsilon=0.5$ for whole ranges for $W$ and $Q^2$ for brevity. 

\begin{figure}[t]
\includegraphics[width=8.5cm]{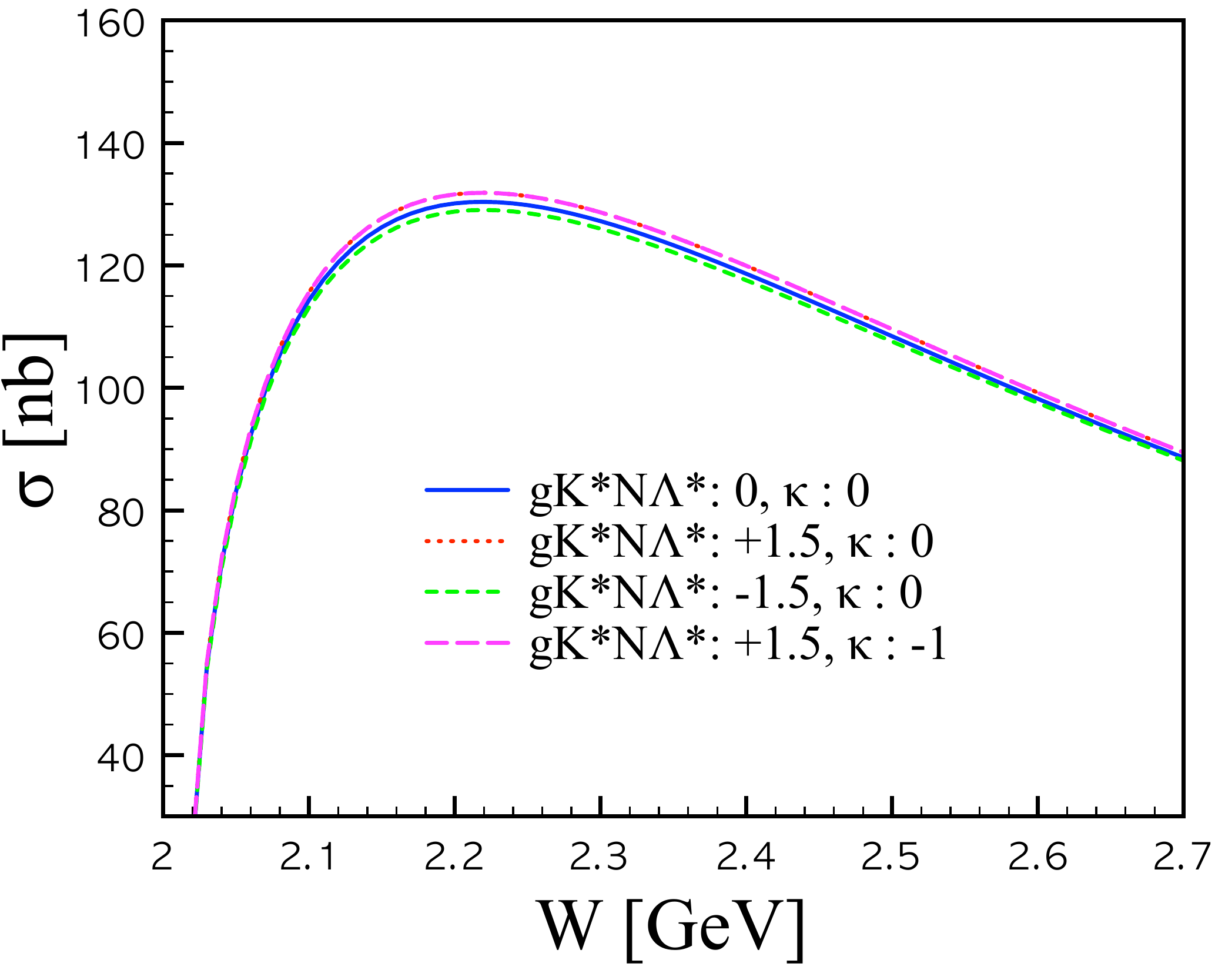}
\caption{(Color online) Total cross section without the resonance contributions, varying the values of $g_{K^*N\Lambda^*}$ and $\kappa_{\Lambda^*}$. Here, we choose the average for $Q^2=(0.9\sim2.4)\,\mathrm{GeV}^2$ and the cutoff mass $\Lambda=1$ GeV in Eq.~(\ref{eq:FORM}).}       
\label{FIG2}
\end{figure}

Before going further, we want to examine the $K^*$-exchange in the $t$ channel and $\kappa_{\Lambda^*}$ effect in the $u$-channel. These contributions were turned out to be negligible in explaining the experimental data for the photoproduction off the proton target~\cite{Nam:2009cv,Nam:2010au}. In Figure~\ref{FIG2}, we plot the total cross sections for the electroproduction off the proton target without the resonance contribution. Here, the photon virtuality is taken to be the center value of the kinematic region $(0.9\sim2.4)\,\mathrm{GeV}^2$ and the cutoff mass to be $1$ GeV as a trial. We examine the various cases of $g_{K^*N\Lambda^*}$ and $\kappa_{\Lambda^*}$: $g_{K^*N\Lambda^*}=(0,\pm1.5)$ and $\kappa_{\Lambda^*}=(0,\kappa_n)$. As shown in the figure, the effects of those contributions are negligible, showing a few percent changes from that without those contributions (solid), although there appears slight increasing in the higher $W$ region with the finite $\kappa_{\Lambda^*}$ value $\sim\kappa_n$. If we take the theoretical estimations from the SU(6) quark model~\cite{Isgur:1978xj,Hyodo:2006uw} and phenomenological study with the $K^*$ Regge trjectory~\cite{Titov:2005kf},  we have $g_{K^*N\Lambda^*}\approx10$ and  $g_{K^*N\Lambda^*}=7.1$ (or $-12.6$), respectively. We verified that these estimations gives only small differences with about $(5\sim10)\%$ in the cross section, comparing to that without the $K^*$ exchange. Hence, taking into account that the tree-level Born approximation is well valid in the low energy region and the contact-term dominance, in addition to the $K$ exchange, as supported theoretically and experimentally, we can rather safely ignore those $K^*$ and $\kappa_{\Lambda^*}$ contributions from the calculations hereafter, i.e. $(g_{K^*N\Lambda^*},\kappa_{\Lambda^*})=0$. 
%FIGURE>>>
%FIGURE<<<

\begin{figure}[t]
\hspace{2cm}
\begin{tabular}{ccc}
\includegraphics[width=6cm]{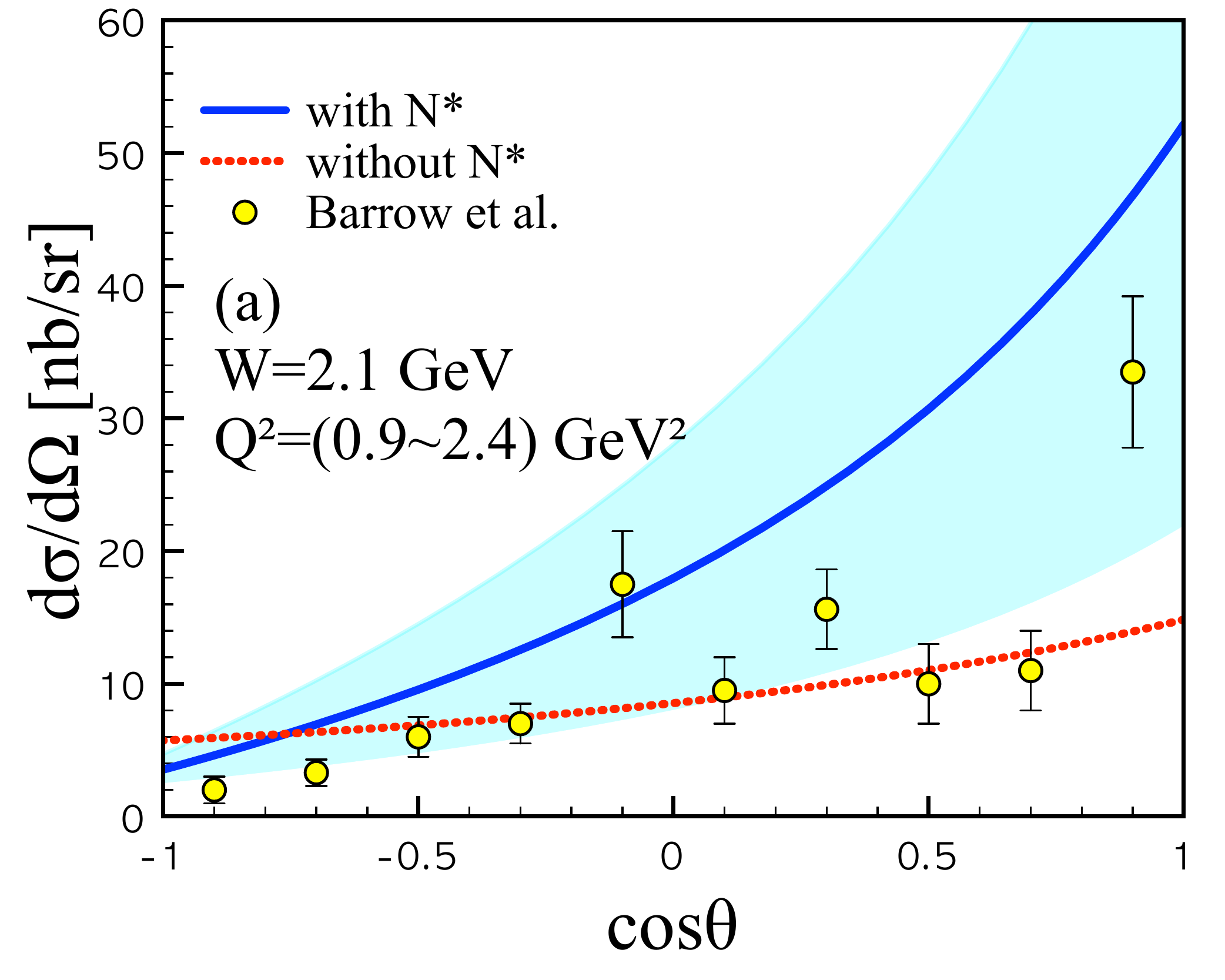}
\includegraphics[width=6cm]{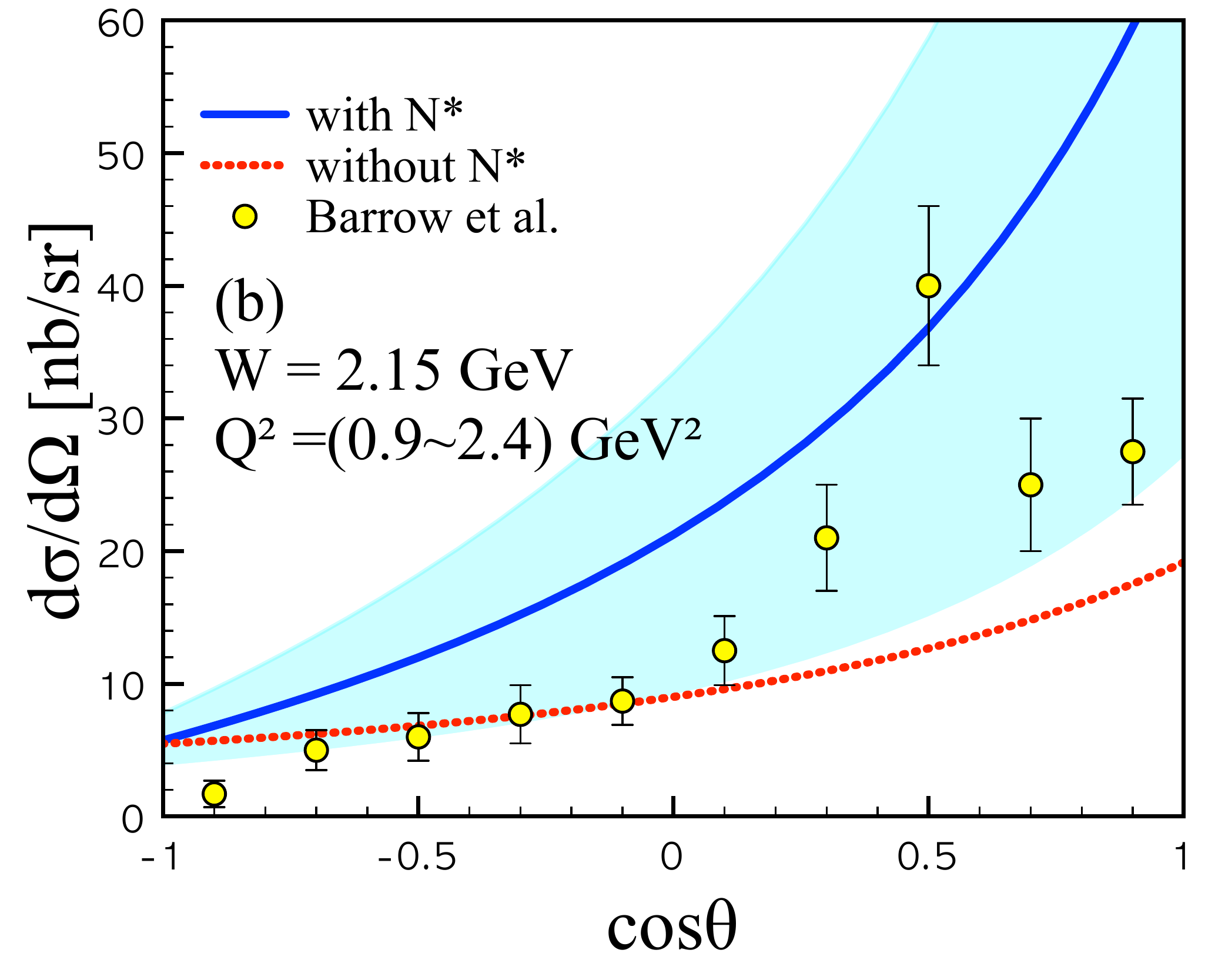}
\includegraphics[width=6cm]{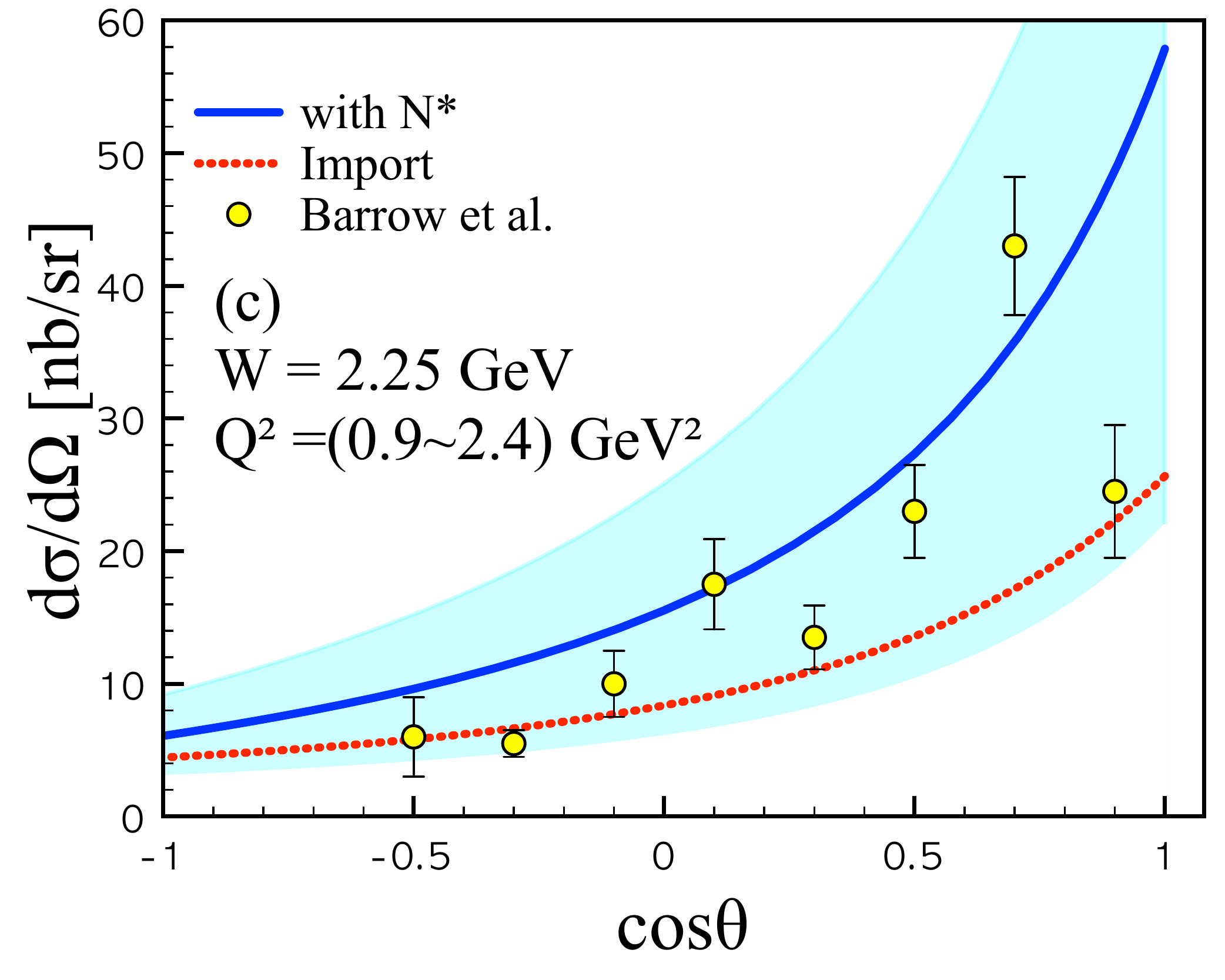}
\end{tabular}
\begin{tabular}{ccc}
\includegraphics[width=6cm]{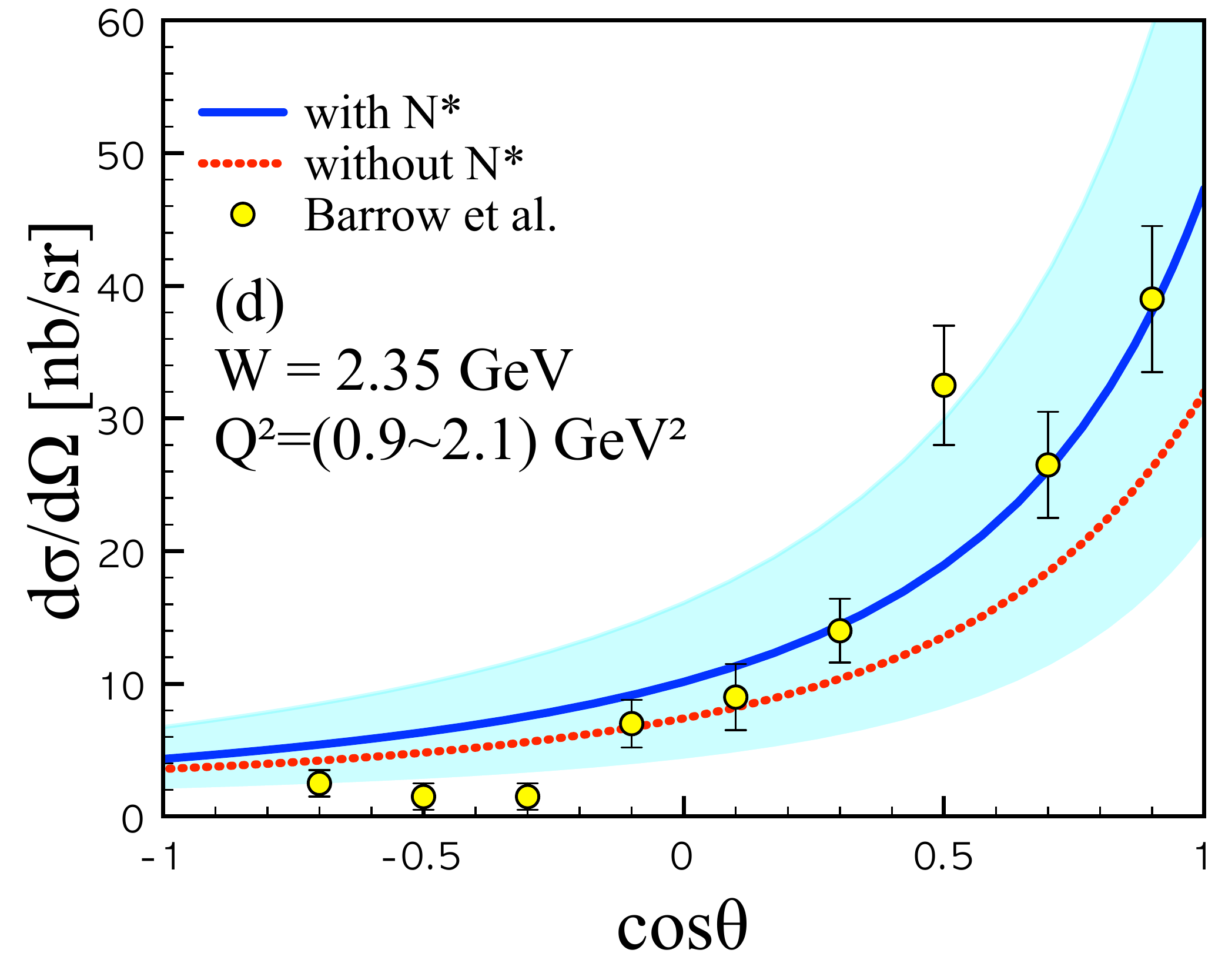}
\includegraphics[width=6cm]{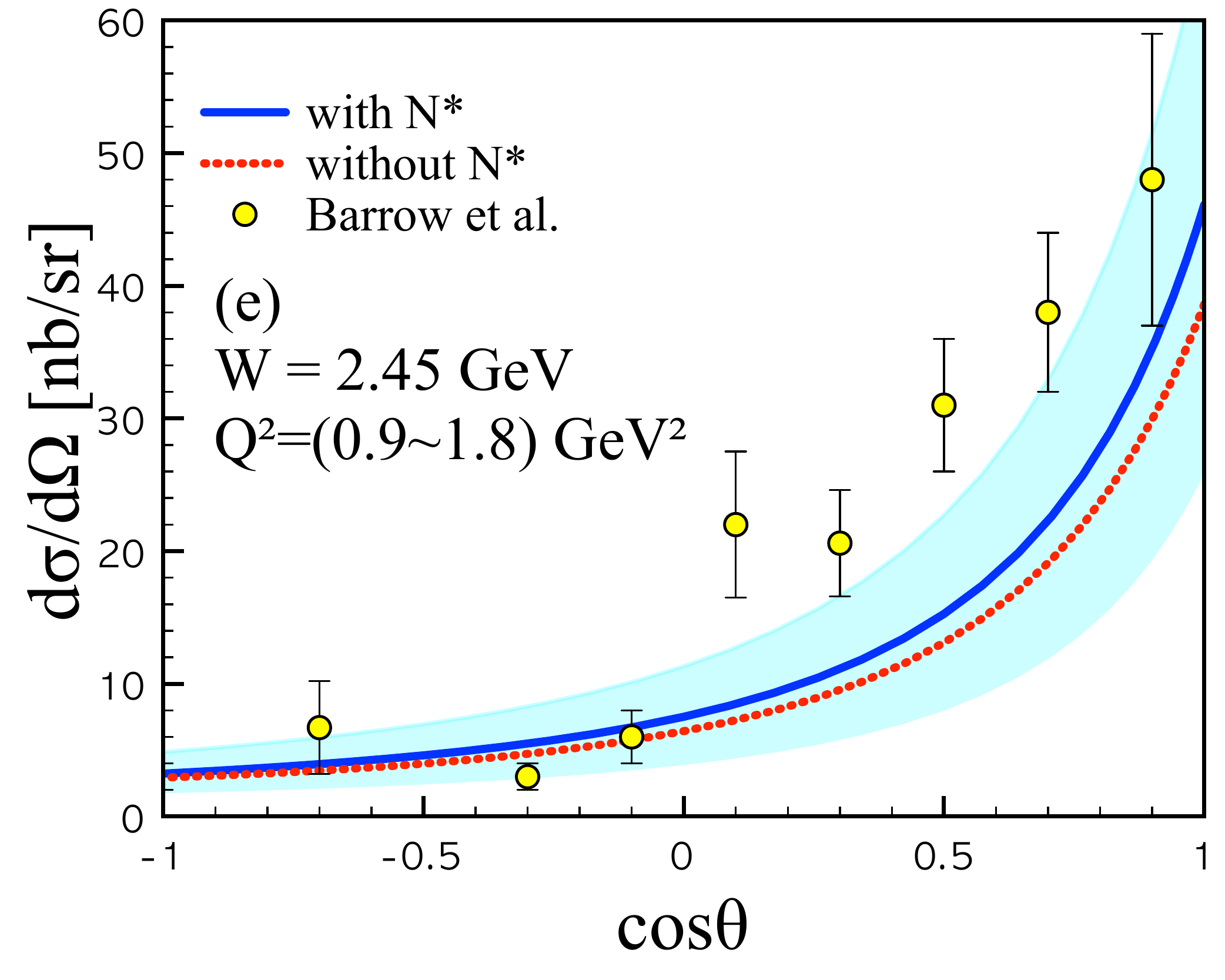}
\includegraphics[width=6cm]{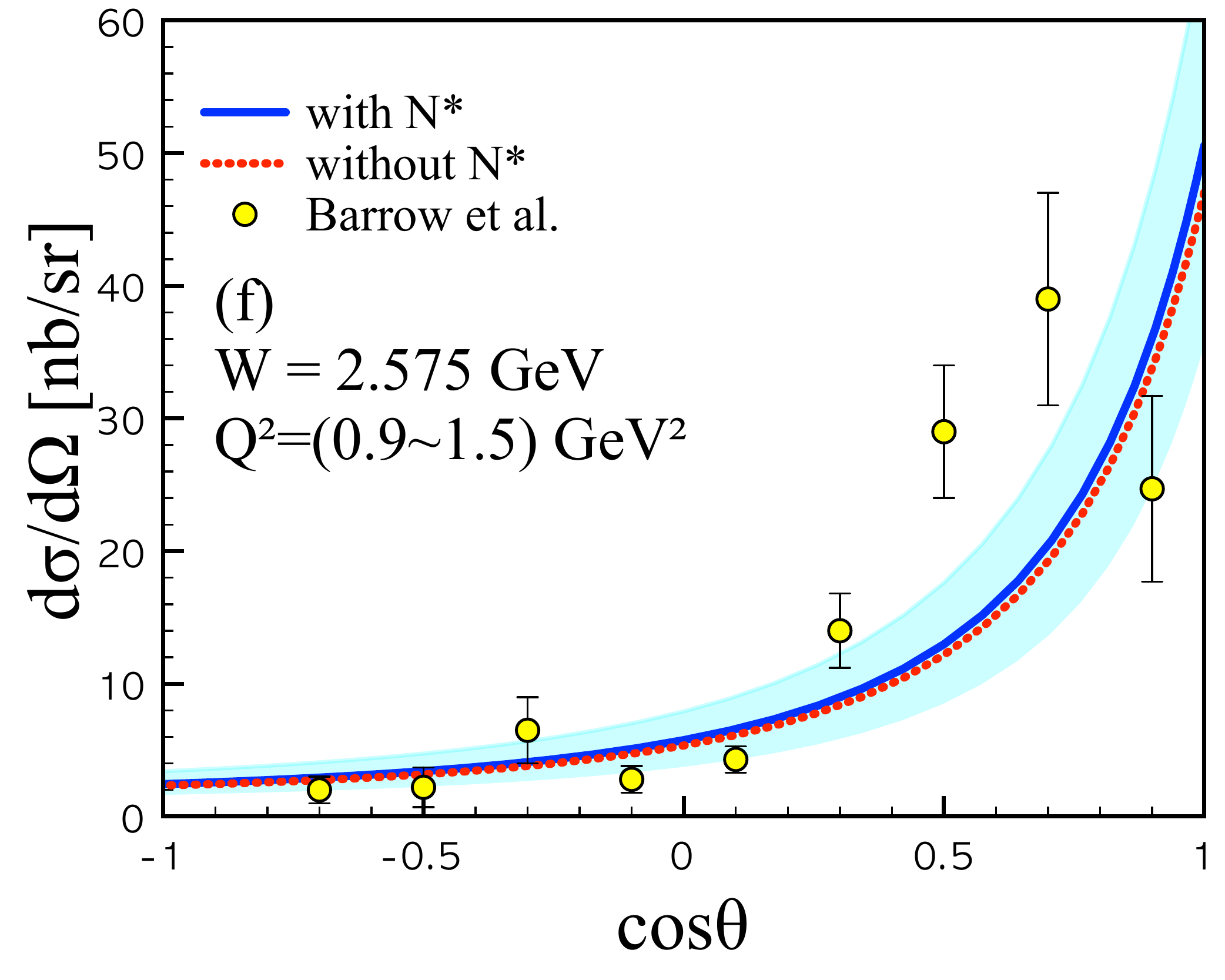}
\end{tabular}
\caption{(Color online) Differential cross section $d\sigma/d\Omega$ [$\mu$b/sr] as functions of $\cos\theta$ for different $W$ values with (solid) and without (dash) the resonance contributions. The shaded area denotes the interval of the photon virtuality, indicated in each panel. Thus, the solid curve are the average for the interval. Experimental data are taken from Ref.~\cite{Barrow:2001ds}. }       
\label{FIG3}
\end{figure}

Considering all the discussions and ingredients mentioned above, we are now ready to compute various physical quantities for the $\Lambda^*$ electroproduction off the proton target. First, we show the numerical results for the differential cross section $d\sigma/d\Omega$ as a function of $\cos\theta$ in Figure~\ref{FIG3}. The curves are the average ones for the photon virtuality interval $Q^2=(0.9\sim2.4)\,\mathrm{GeV}^2$ for $W=(2.10,2.15,2.25,2.35,2.45,2.57)$ in the panel $(a\sim f)$.  Here, we fix the cutoff masses for the form factors in Eqs.~(\ref{eq:fc}) and (\ref{eq:DPFF}) to reproduce the data~\cite{Barrow:2001ds} as follows:
%EQUATION>>>
\begin{equation}
\label{eq:CUTOFF}
\Lambda_h=1000\,\mathrm{MeV},\,\,\,\,
\Lambda_{N,\mathrm{EM}}=950\,\mathrm{MeV},\,\,\,\,
\Lambda_{K,\mathrm{EM}}=1050\,\mathrm{MeV}.
\end{equation}
%EQUAITON<<<
At the same time, the phase factor for the resonances are given as $(\phi_1,\phi_3,\phi_5)=(-\pi/2,\pi/2,\pi/2)$ in Eq.~(\ref{eq:AMP}).  Moreover, the full decay width for the nucleon resonances are chosen to be $\Gamma_{1,3,5}\approx300$ MeV for the electroproduction of $\Lambda^*$, since this value is close to their average values. We will use these values for the numerical calculations for all the electroproduction of $\Lambda^*$ hereafter. The solid and dot curves represent with and without the resonance contributions. The shaded area indicates the possible region for the photon-virtuality range. From the figure, one can see that the resonance contribution plays an important role to reproduce the data with considerably good agreement. As for $W\lesssim2.3$ GeV, the resonance contributions represent obvious improvements to the results, while the differences between the curves with and without it get diminished as $W$ increases. Note that the shaded areas for the photon-virtuality range cover obviously  the experimental data for all the $V$ values. Note that this sizable resonance contribution is quite different from the conclusion of Ref.~\cite{Barrow:2001ds}, in which the nucleon resonance contributions are assumed to be inappreciable. 
%FIGURE>>>
%FIGURE<<<

\begin{figure}[t]
\begin{tabular}{cc}
\includegraphics[width=8.5cm]{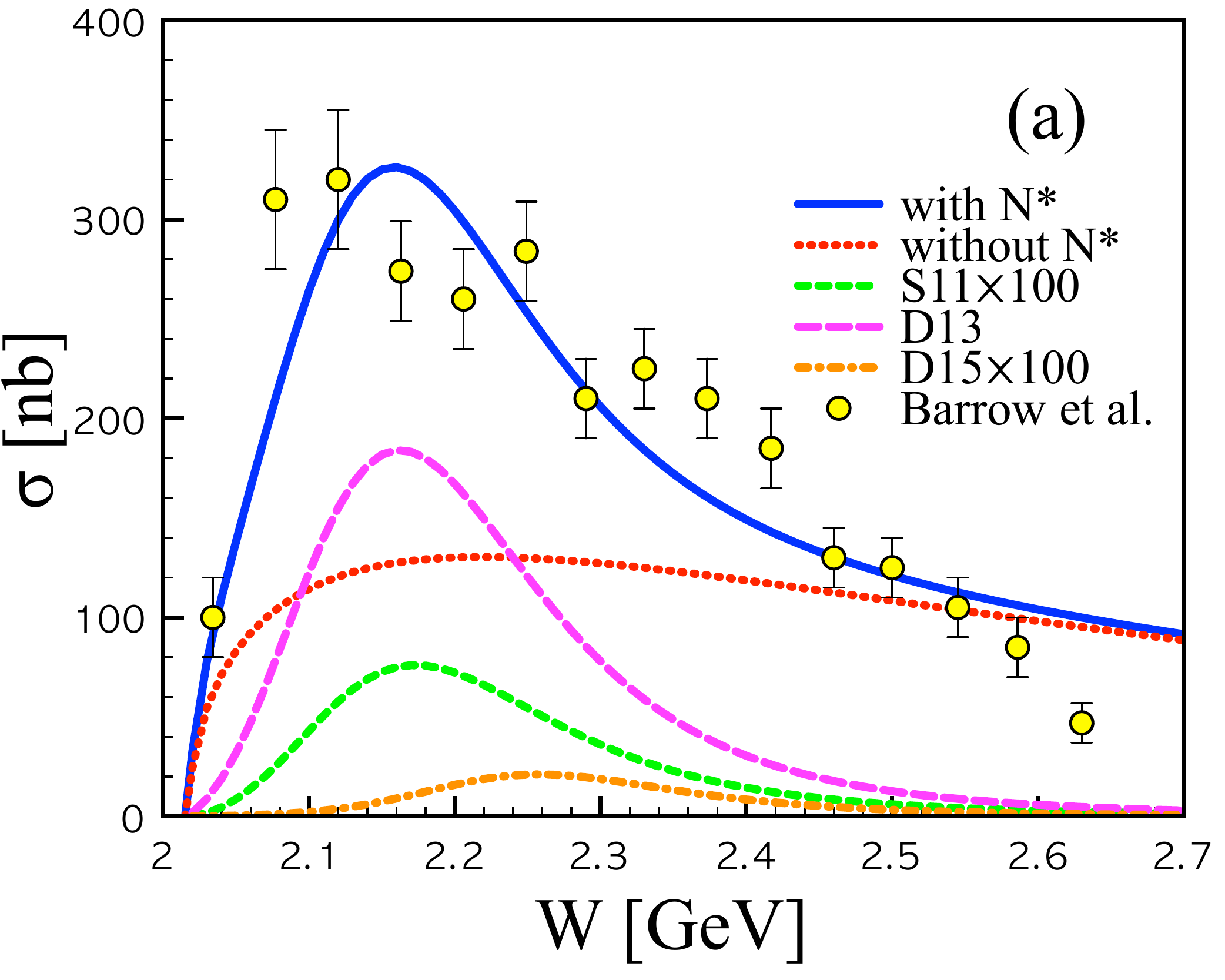}
\includegraphics[width=8.5cm]{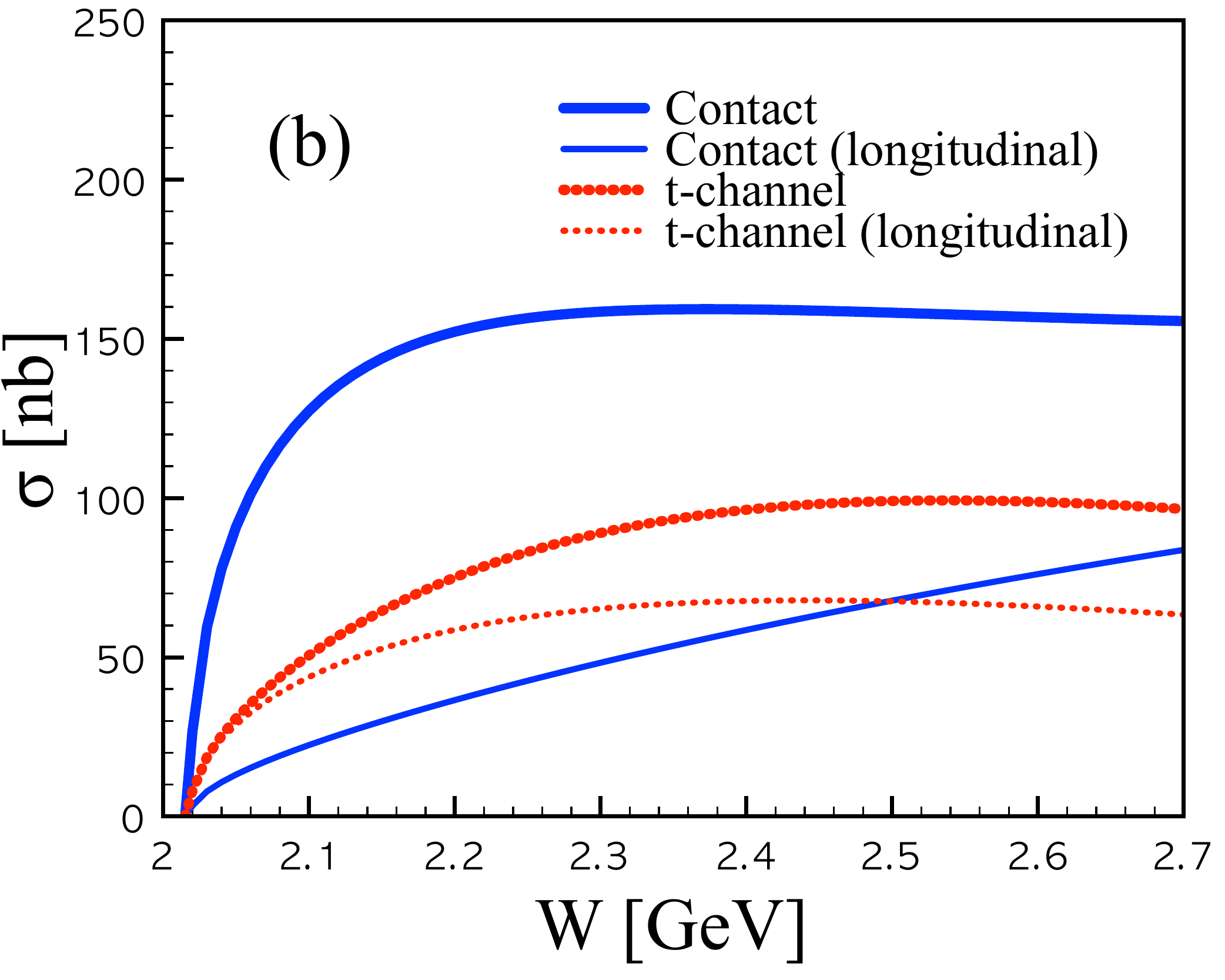}
\end{tabular}
\begin{tabular}{cc}
\includegraphics[width=8.5cm]{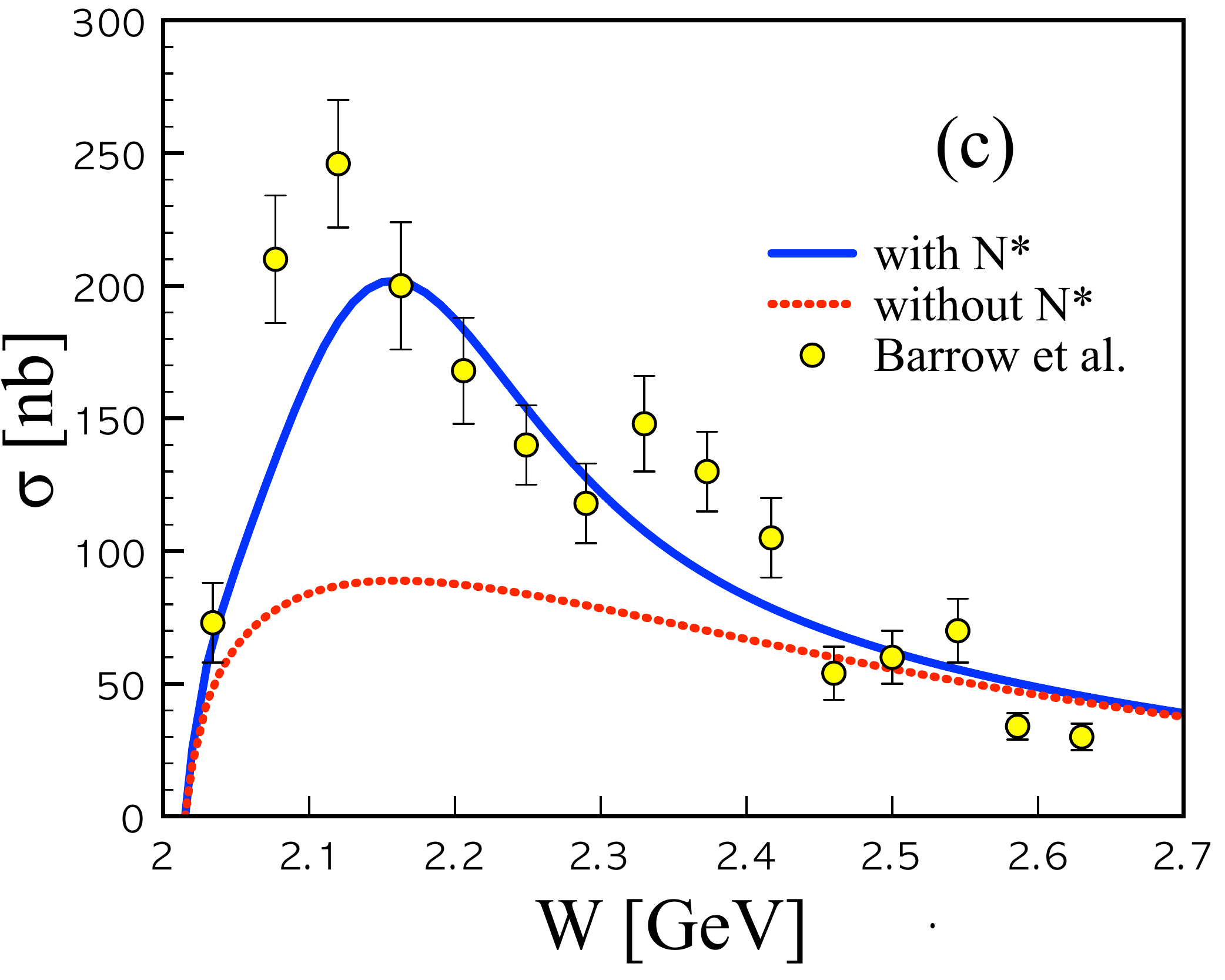}
\includegraphics[width=8.5cm]{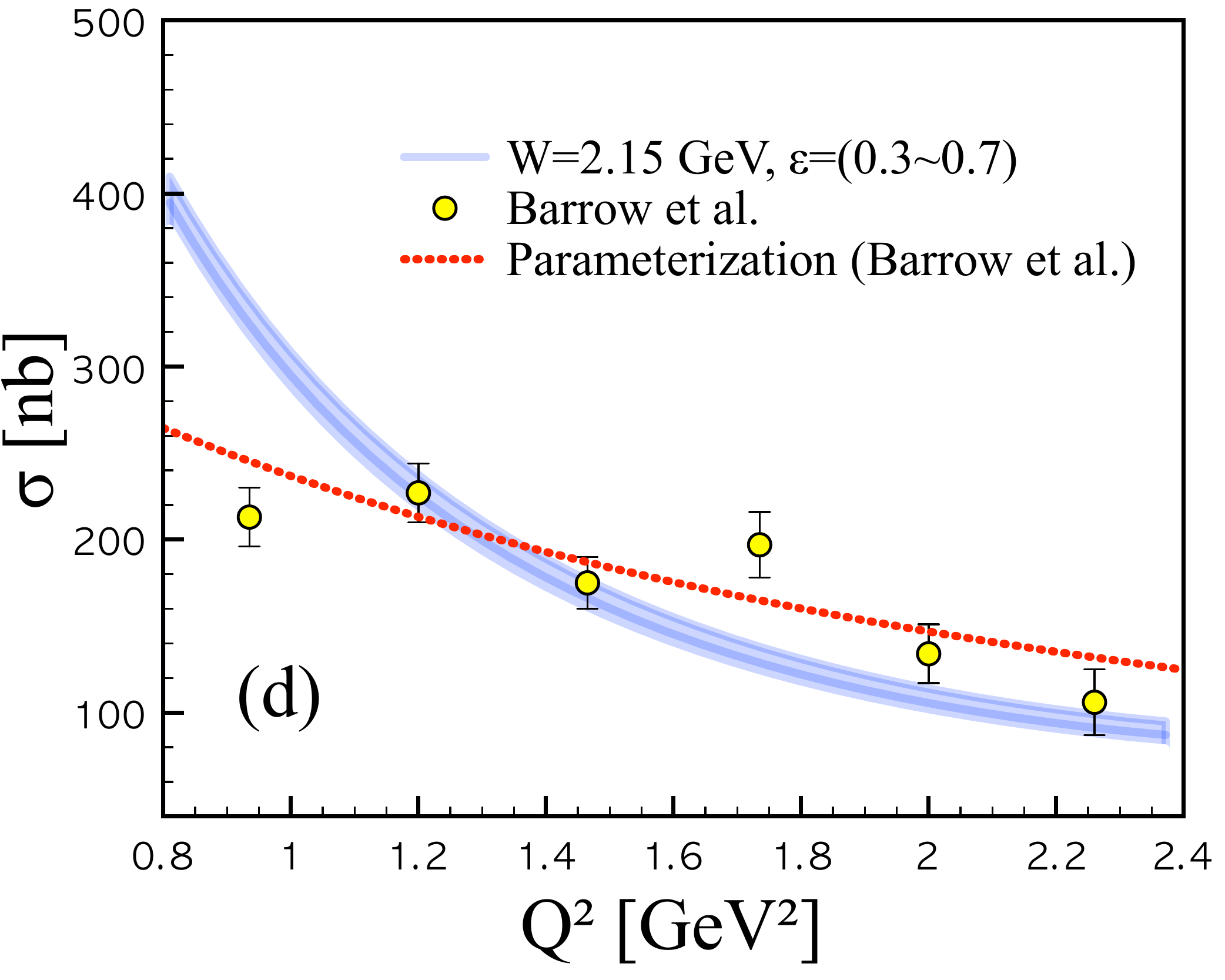}
\end{tabular}
\caption{(Color online) (a) Total cross sections for  $-1\le\cos\theta\le1$ as functions of $W$ with $Q^2=(0.9\sim2.4)\,\mathrm{GeV}^2$. The solid and dash line denote those with and without the resonance contributions, respectively, whereas the shaded area denote the photon virtuality interval. (b) Total cross sections from the contact-term and $t$-channel $K$-exchange contributions for the transverse and longitudinal photons,, separately. (c) Total cross sections  $\cos\theta\le0.6$ ($\theta\gtrsim 53^\circ$) as functions of $W$ in the same manner with (a). (d) Total cross section as a functions of $Q^2$ for $W=2.24$ GeV (solid) and $2.34$ GeV (dot). The dashed line stands for the parameterized curve by $\sigma\propto(m^2_\mathrm{para}+Q^2)^{-2}$ with $m^2_\mathrm{para}=2.73\,\mathrm{GeV}^2$~\cite{Barrow:2001ds}. All the experimental data are taken from Ref.~\cite{Barrow:2001ds}}       
\label{FIG4}
\end{figure}

In the panel (a) of Figure~\ref{FIG4}, we present the numerical results for the total cross sections as functions of $W$ with (solid) and without (dot) the resonance contributions for the whole $K^+$ angle regions: $-1\le\cos\theta\le1$. We also show the resonance contributions separately for the $S_{11}$ (dashed), $D_{13}$ (long-dashed), and $D_{15}$ (dot-dahsed) contributions. The experimental data are taken from Ref.~\cite{Barrow:2001ds}. We observe that $D_{13}$  plays a crucial role to reproduce the data in the vicinity of $W=2.1$ GeV. In the panel (b) of Figure~\ref{FIG4}, we depict the total cross sections from the contact-term (solid) and $K$-exchange (dash) contributions separately for the whole polarization states averaged (thick) and the longitudinal one only (thin), since we verified that these two channels dominates the production process beside the resonance contributions. As understood by each curves, the longitudinal-polarization contribution effects much on the $K$ exchange in the $t$ channel, in comparison to the contact-term one as expected, i.e. the longitudinal component selects the $K$ exchange. Numerically, about $70\%$ of the total production rate is produced from the longitudinal polarization for the $K$ exchange contribution, whereas only about $30\%$ from it for the contact-term one. Note that this observation is quite different from the photoproduction case, in which the contact-term contribution almost dominates the production rate. This interesting tendency can be understood in detail by the following: In our kinematical setup, $k_3\cdot\epsilon$ in the $K$ exchange in the $t$ channel in Eq.~(\ref{eq:AMP}) can be written by
%EQUATION>>>
\begin{equation}
\label{eq:PRO}
k_3\cdot\epsilon_x=\mathrm{k}_3\sin\theta,\,\,\,\,
k_3\cdot\epsilon_y=0,\,\,\,\,
k_3\cdot\epsilon_z=\frac{\sqrt{2\varepsilon^*}}{|\bm{q}|}\left[\mathrm{k}_1E_3-\mathrm{k}_3E_1\cos\theta \right],
\end{equation}
%EQUAITON<<<
where $\mathrm{k}_{1,3}$ denote the three momenta for $k_{1,3}$. Note that the last term, $k_3\cdot\epsilon_z$ only exists for the electroproduction case and enhances the production rate  by $(k_3\cdot\epsilon_z)^2$ in comparison to the photoproduction. On  the contrary, the contact term, which contains $(\epsilon_{\Lambda^*}\cdot\epsilon)$, in which $\epsilon_{\Lambda^*}$ stands for the vector part of the Rarita-Schwinger field, is not much affected by the longitudinal component of the virtual photon. These differences between the two dominant contributions for the electroproduction indicate that the $\phi$ distribution in the GJ frame can be different from that of the photoproduction, in which the spin-$3/2$ states of $\Lambda^*$ decay into $K^-p$ mainly via the contact-term contribution. We will examine this difference in detail below soon. In the panel (c) of Figure~\ref{FIG4}, we also draw the numerical results for the total cross section in the same manner with the panel (a) for the limited $\theta$-angle region, i.e. $\cos\theta\le0.6$, comparing with the experimental data~\cite{Barrow:2001ds}. In the presence of the $D_{13}(2150)$ contribution, we can reproduce the experimental data qualitatively well, while the production rate for the region $W=(1.95\sim2.2)$ is still  underestimated. In the panel (d) of Figure~\ref{FIG4}, we depict the numerical results for the total cross sections as functions of $Q^2$ for $W=2.15$ GeV for $\varepsilon=(0.3\sim0.7)$, represented by the shade area. Experimental data and the parameterized curve (dot) are taken from Ref.~\cite{Barrow:2001ds}. The parameterization was done by $\sigma\propto(m^2_\mathrm{para}+Q^2)^{-2}$ with $m^2_\mathrm{para}=2.73\,\mathrm{GeV}^2$. We observe that the experimental data are qualitatively well reproduced within the model, considering the experimentally accessible photon virtuality $Q^2=(0.9\sim2.4)\,\mathrm{GeV}^2$, although there appears considerable overshoot in the small $Q^2$ region. This overshoot can be cured by employing more realistic $Q^2$ dependence for nucleon resonances. However, we will not cover this interesting issue here and leave it for the future works. 
%FIGURE>>>
%FIGURE<<<

\begin{figure}[t]
\includegraphics[width=8.5cm]{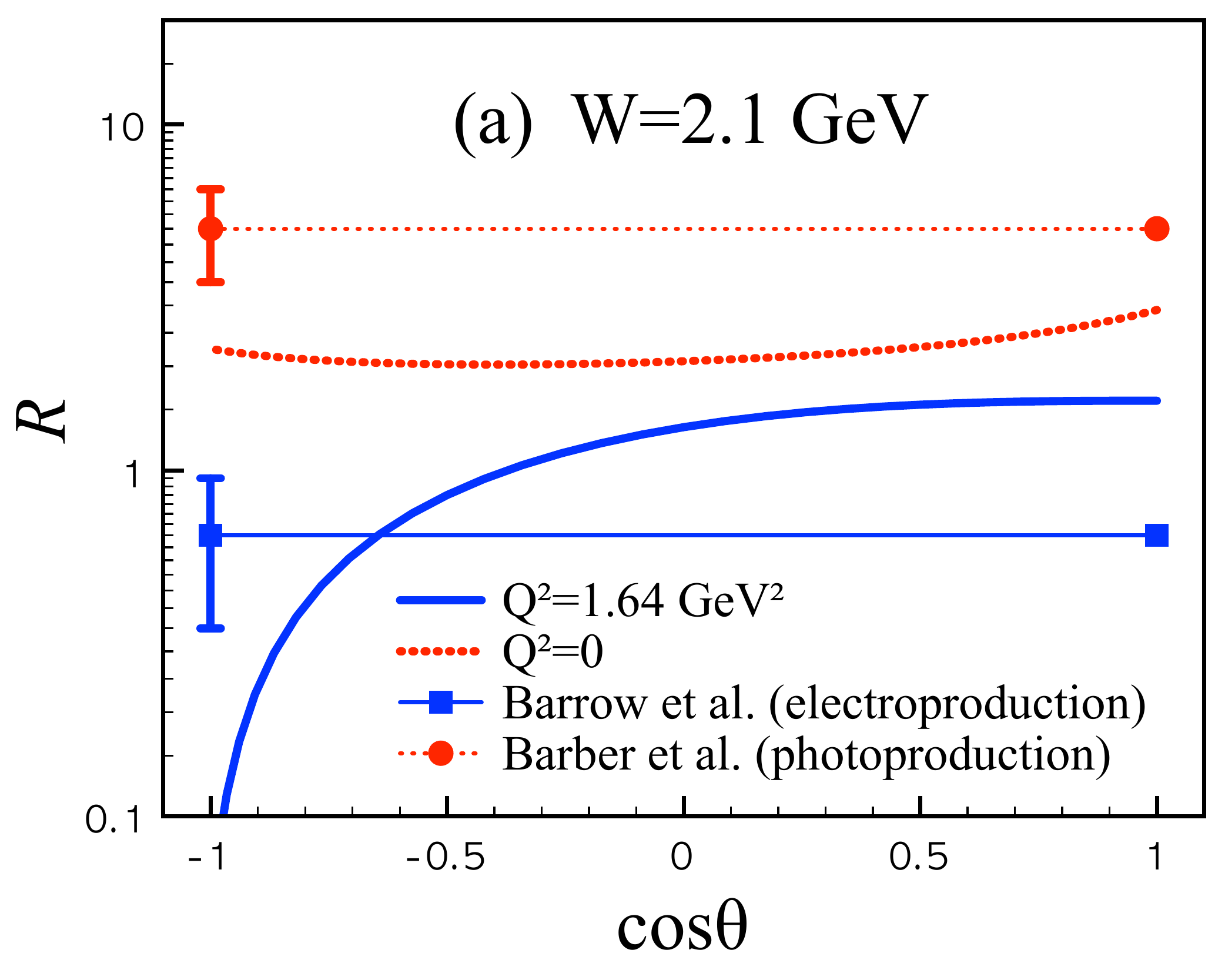}
\caption{(Color online) Ratio for the $\Lambda(1520)$ spin-state distribution, $\mathcal{R}$ in Eq.~(\ref{eq:RATIO}) as functions of $\cos\theta$ for $Q^2=0$ (solid) and $1.64\,\mathrm{GeV}^2$ (dash) at $W=2.1$ GeV. Experimental data are taken from the LAMP2~\cite{Barber:1980zv} (circle) and CLAS~\cite{Barrow:2001ds} (square) collaborations.}       
\label{FIG5}
\end{figure}

From now on, we want to discuss the $\phi$ distribution in the GJ frame as discussed in several literatures~\cite{Barber:1980zv,Muramatsu:2009zp,Nam:2010au}. Considering the $\Lambda^*$-rest frame, one can construct a kinematic frame with the colliding meson and target nucleon, producing $\Lambda^*$ at rest, then it decays into $K^-$ and $p$, i.e. the GJ frame. Interestingly, the angular distribution of $K^-$ can be fully derived from the simple spin statistics of the system~\cite{Barber:1980zv}. The $\phi$ distribution can be simply written as follows: 
%EQUATION>>>
\begin{equation}
\label{eq:DF}
\mathcal{F}_{K^-}(\phi)
\approx \left[A\sin^{2}\phi \right]_{S_{\Lambda^*}=3/2}
+\left[B\left(\frac{1}{3}+\cos^{2}\phi\right) \right]_{S_{\Lambda^*}=1/2},
\end{equation}
%EQUAITON<<<
where $\phi$ denotes the decaying angle of $K^-$ in the $\Lambda^*$ rest frame. If the $K^-$ decays from $\Lambda^*(S=3/2)$, the distributions follows the first term in the right-hand-side of Eq.~(\ref{eq:DF}). On the contrary, if it is does from $\Lambda^*(S=1/2)$ state, it can be described by the second term. Here, $A$ and $B$ are real values and stand for the relative strength for each spin states, satisfying the normalization $A+B\approx1$. In Ref.~\cite{Barrow:2001ds}, the authors included an additional term $C\cos\phi$, indicating the interferences with the backgrounds, although we omit it for simplicity here. Theoretically, the strength factors $A$ and $B$ can be estimated and computed by the following parameterization, remembering that the final-state $\Lambda^*$ spins are just summed,
%EQUATION>>>
\begin{equation}
\label{eq:AAA}
A=\frac{d\sigma_{\Lambda^*(S=3/2)}}{d\sigma_\mathrm{total}},\,\,\,\,
B=\frac{d\sigma_{\Lambda^*(S=1/2)}}{d\sigma_\mathrm{total}},
\end{equation}
%EQUAITON<<<
satisfying the normalization condition~\cite{Nam:2010au}. We also define the ratio of the coefficients for further purpose by
%EQUATION>>>
\begin{equation}
\label{eq:RATIO}
\mathcal{R}=\frac{A}{B}.
\end{equation}
%EQUAITON<<<
If $\mathcal{R}$ is (larger, smaller) than unity, the $\Lambda^*(S=3/2,1/2)$ decay will dominates the process. If one draws the curves of the $\phi$ distributions in Eq.~(\ref{eq:DF}) for $\Lambda^*(S=3/2)$, the curve shape becomes a hill ($\frown$), whereas a valley ($\smile$) for $\Lambda^*(S=1/2)$, as functions of $\cos\phi$. For convenience, we assign the first by {\it A-type} and the second by {\it B-type}, considering their coefficients named. In Figure~\ref{FIG5}, we draw the ratio $\mathcal{R}$ in Eq.~(\ref{eq:RATIO}) as functions of $\cos\theta$ for $Q^2=0$ (solid), i.e. photoproduction, and $Q^2=1.64\,\mathrm{GeV}^2$ representing the (almost) average value for the CLAS experiment~\cite{Barrow:2001ds}. here, we take $W=2.1$ GeV. The horizontal solid and dot lines stand for the experimental data taken from the  CLAS (Barrow {\it et al.}) (square)~\cite{Barrow:2001ds} and the LAMP2 (Barber {\it et al.})~\cite{Barber:1980zv} (circle) data, with the errors. It turns out that the theoretical curves show qualitative similarities with the experimental data: This observation indicates theoretically that the contact-term dominate the photoproduction of $\Lambda^*$, whereas the contact term and $K$ exchange play similar roles for the electroproduction. 
%FIGURE<<<

\begin{figure}[t]
\begin{tabular}{cc}
\includegraphics[width=8.7cm]{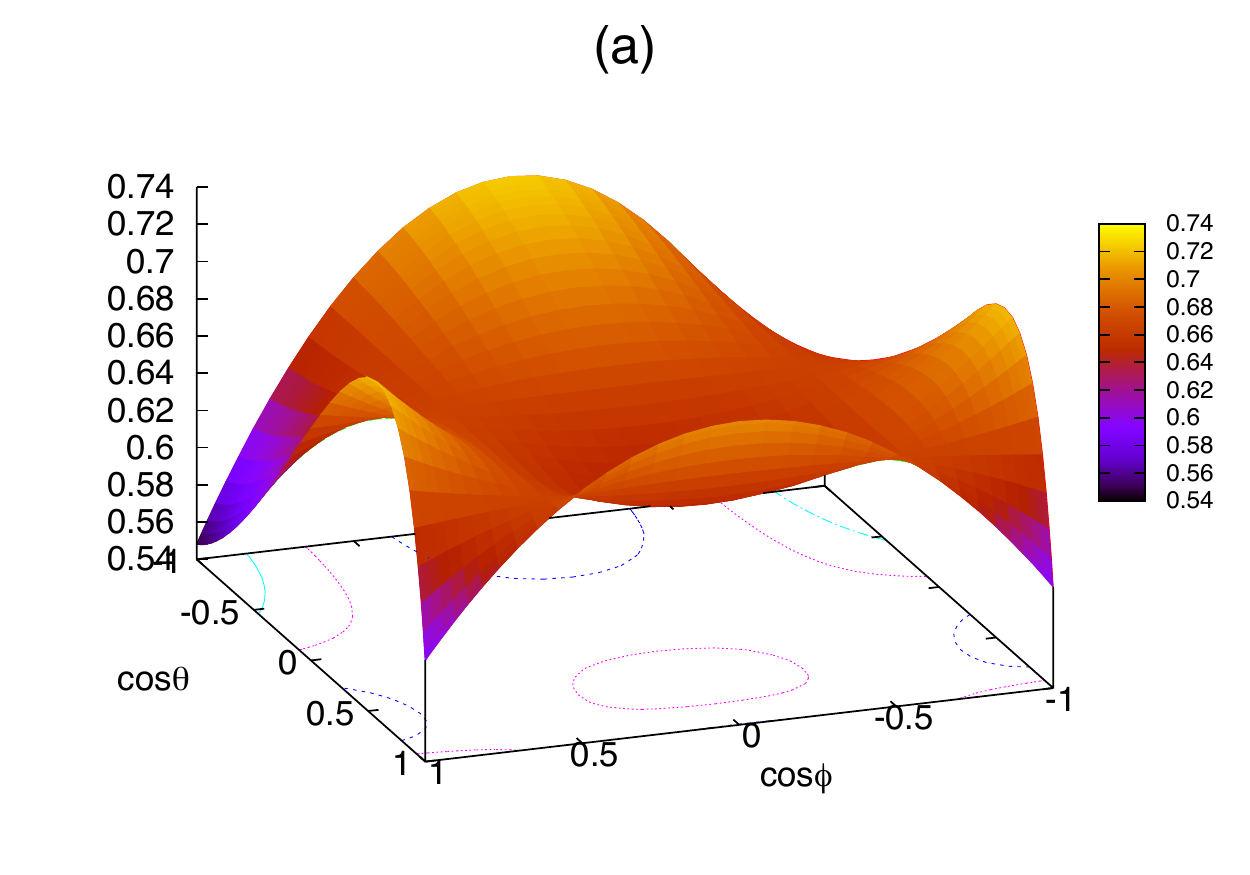}
\includegraphics[width=8.7cm]{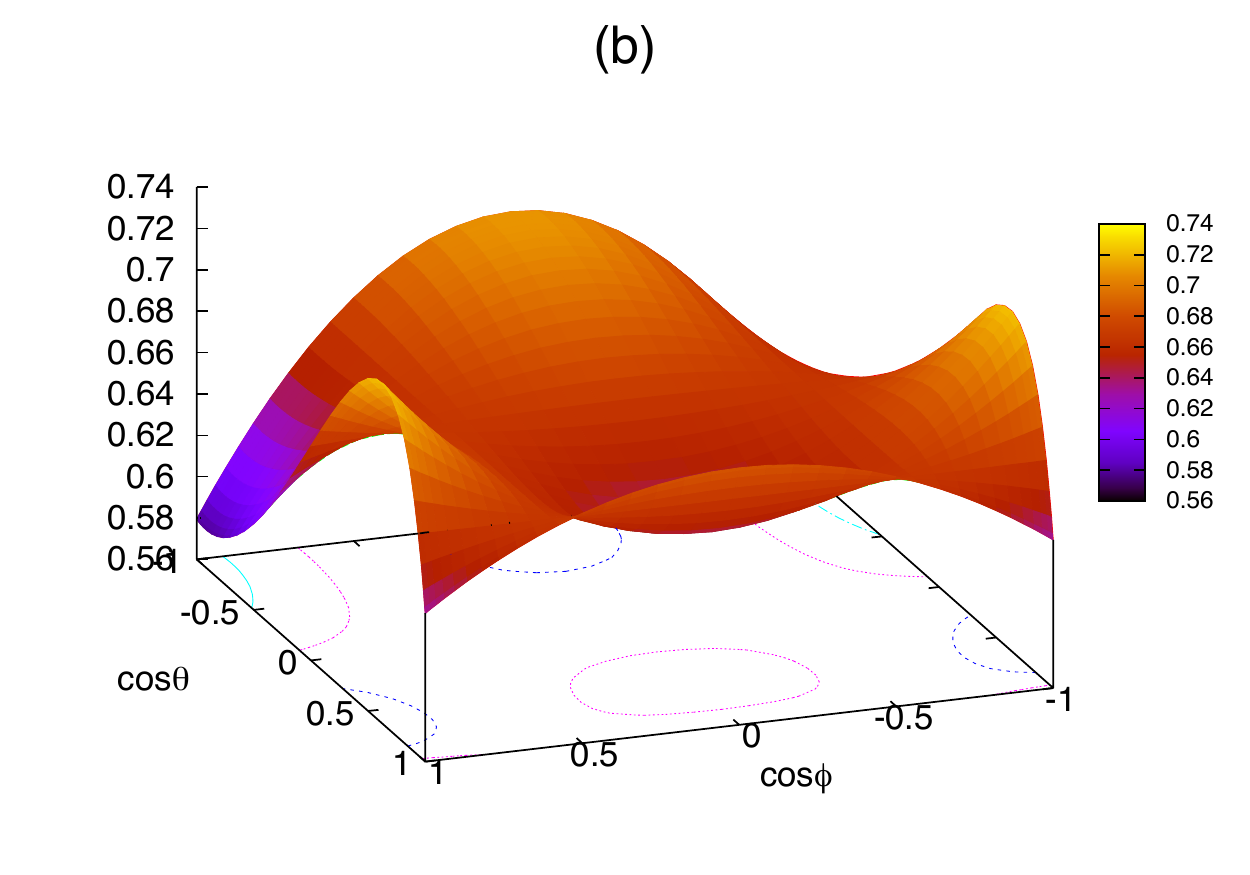}
\end{tabular}
\begin{tabular}{cc}
\includegraphics[width=8.7cm]{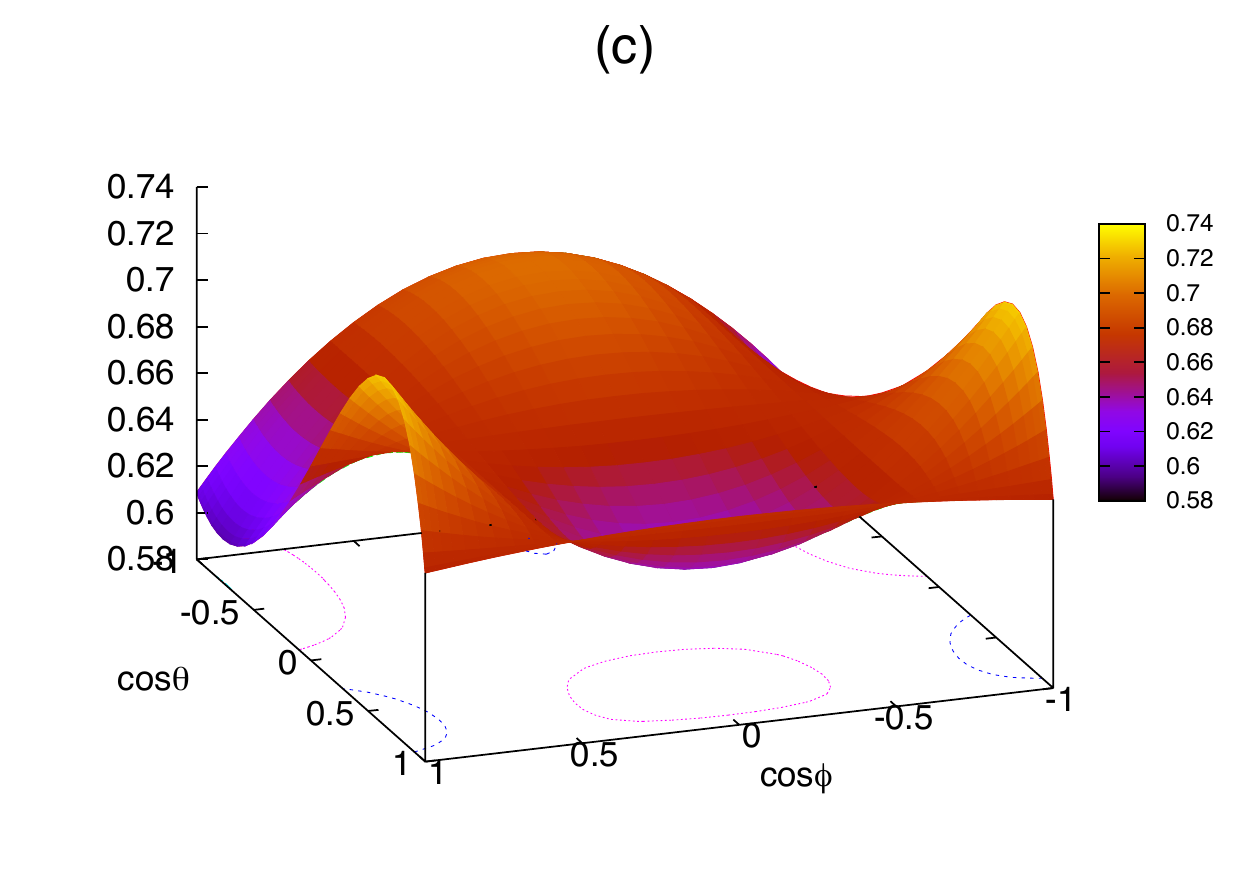}
\includegraphics[width=8.7cm]{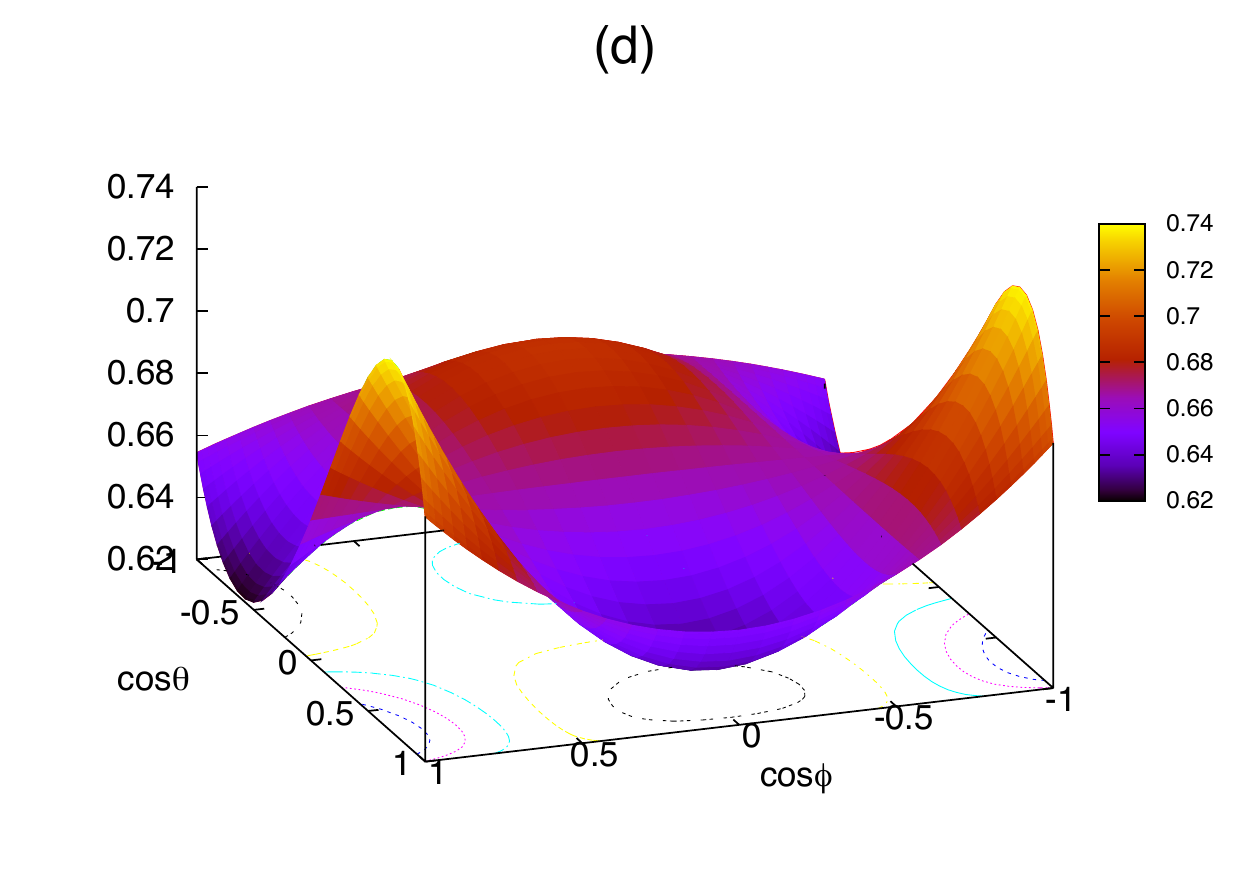}
\end{tabular}
\includegraphics[width=8.5cm]{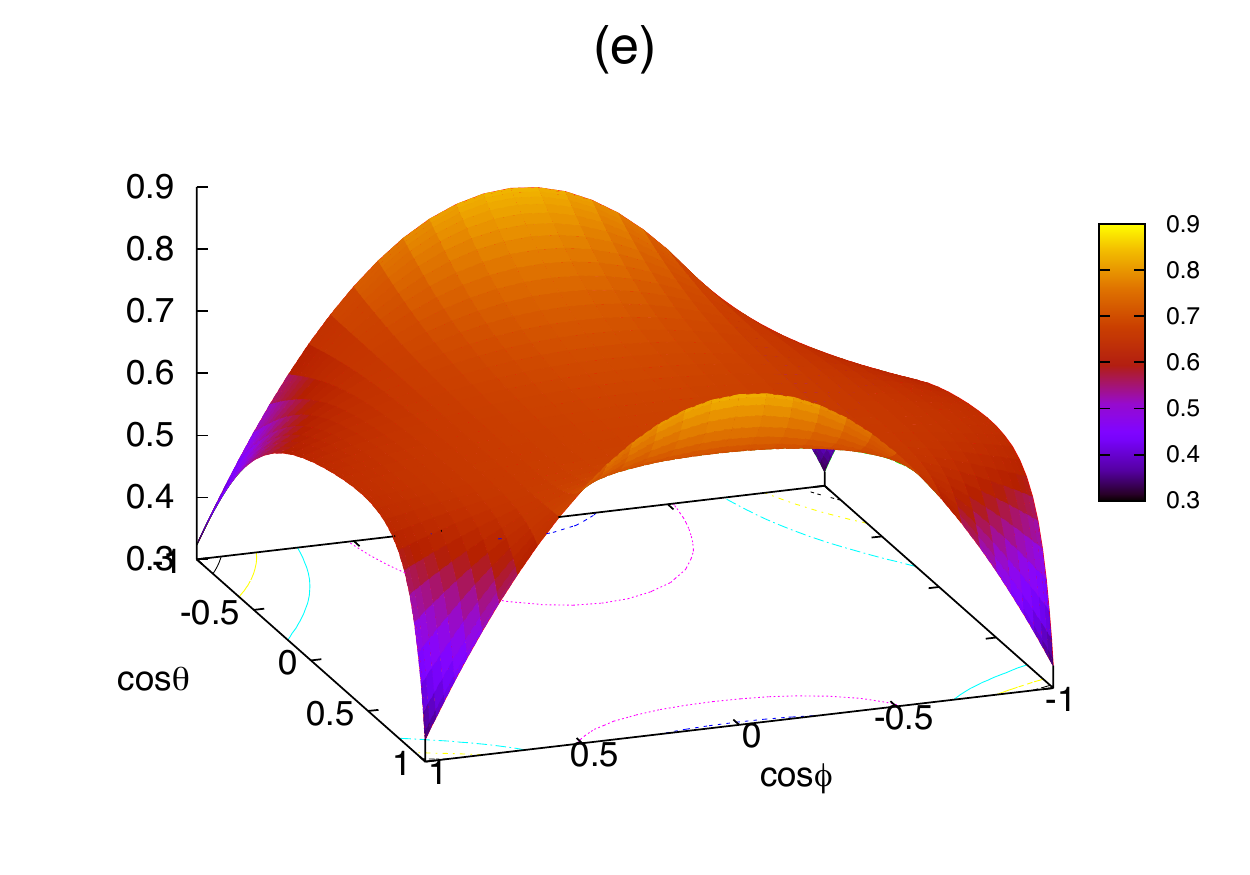}
\caption{(Color online) $K^-$ decay angle ($\phi$) distribution in Eq.~(\ref{eq:DF}) for the $\Lambda^*$ electroproduction as functions of $\cos\phi$ and $\cos\theta$ at $W=2.4$ GeV, for $Q^2=1.05\,\mathrm{GeV}^2$ (a), $1.35\,\mathrm{GeV}^2$ (b), $1.65\,\mathrm{GeV}^2$ (c), and $2.10\,\mathrm{GeV}^2$ (d).  In panel (e), we plot the same for the photoproduction, i.e. $Q^2=0$. For all the cases, we included the resonance contributions.}       
\label{FIG6}
\end{figure}

\begin{figure}[t]
\includegraphics[width=8.5cm]{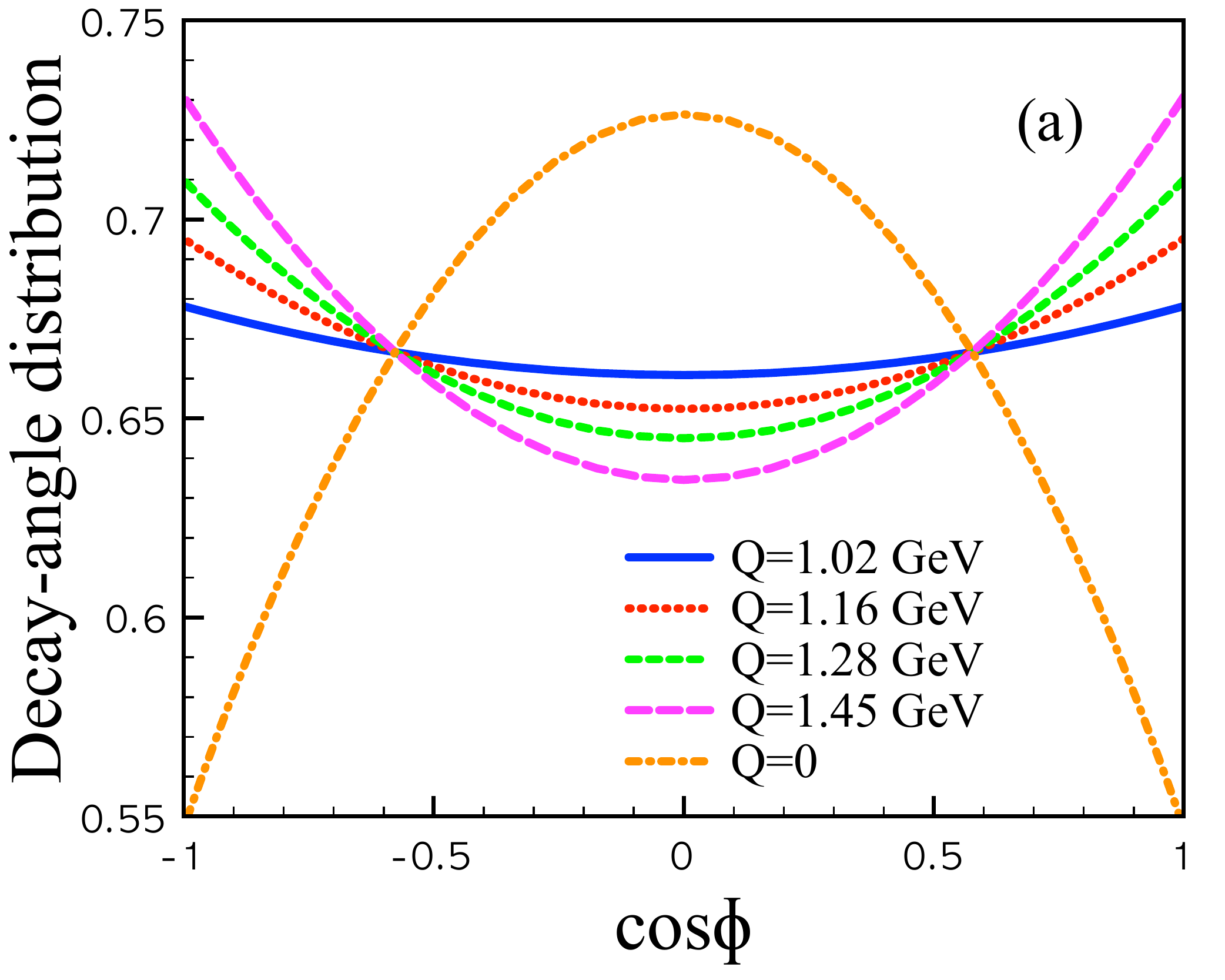}
\caption{(Color online) $K^-$ decay angle ($\phi$) distribution in Eq.~(\ref{eq:DF}) for the $\Lambda^*$ electroproduction as a function of $\cos\phi$ at $\theta=30^\circ$ and $W=2.4$ GeV for $Q^2=(1.02,1.16,1.28,1.45)\,\mathrm{GeV}^2$. The experimental data are taken from Ref.~\cite{Barrow:2001ds}. }       
\label{FIG7}
\vspace{1cm}
\begin{tabular}{ccc}
\includegraphics[width=6.0cm]{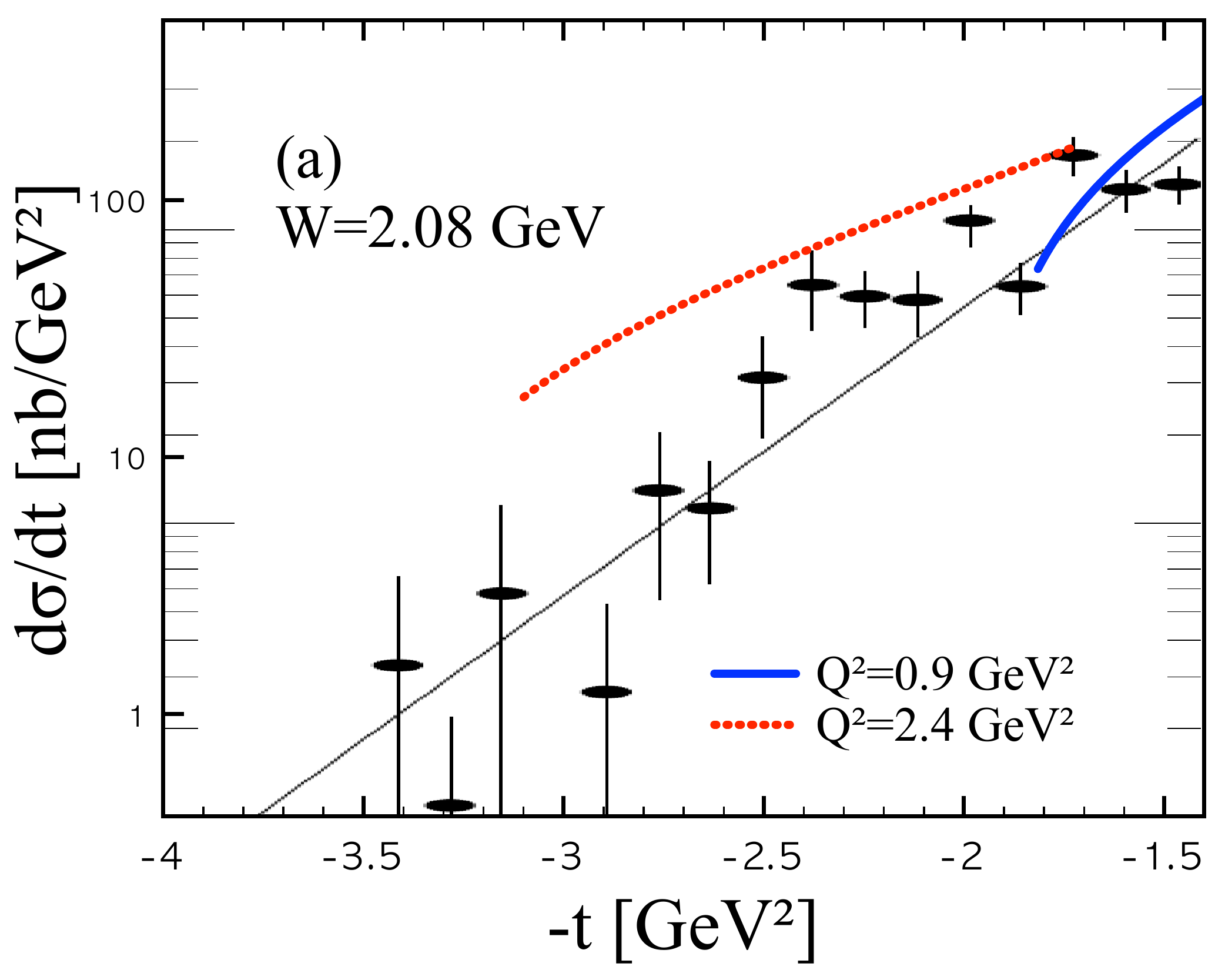}
\includegraphics[width=6.0cm]{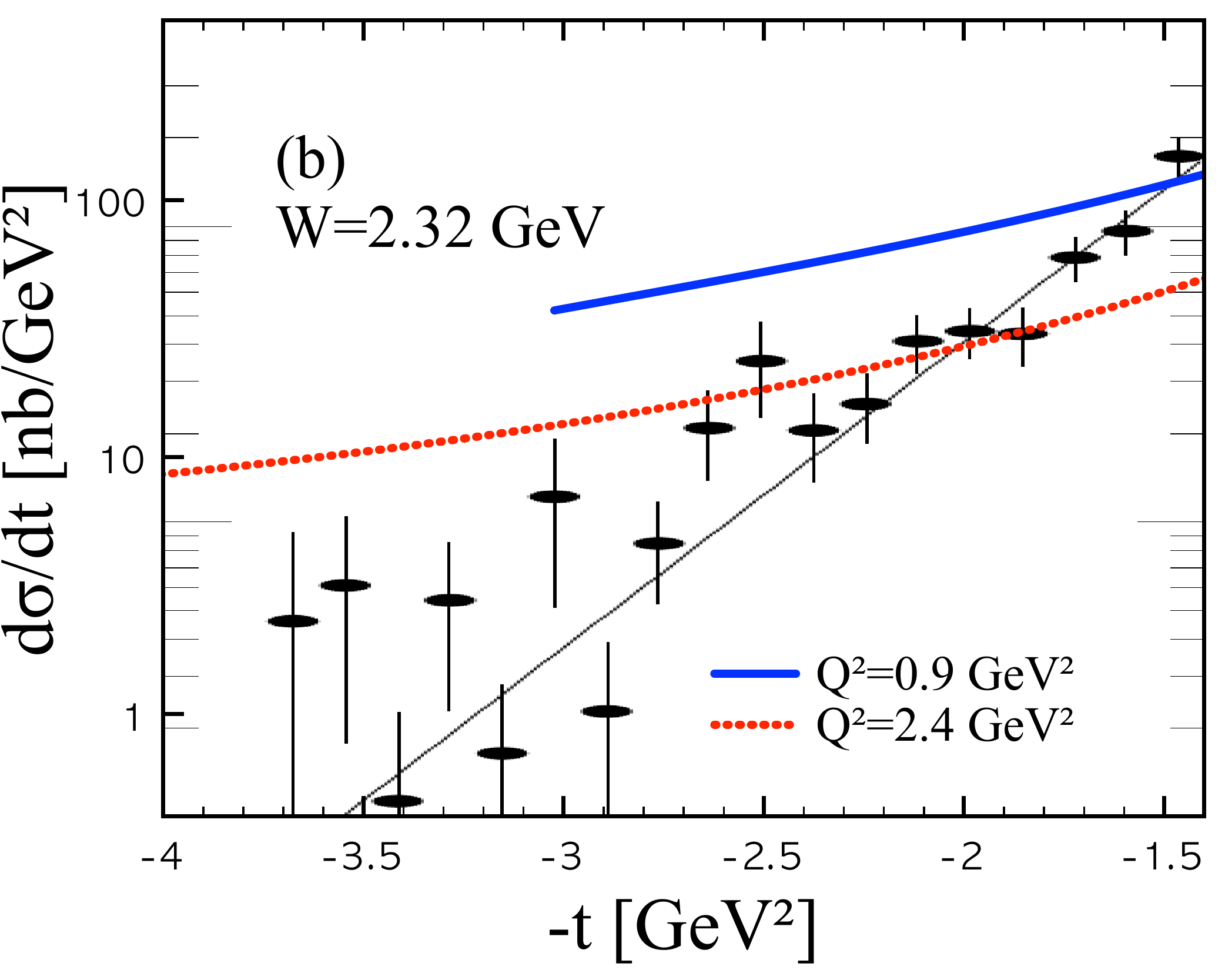}
\includegraphics[width=6.0cm]{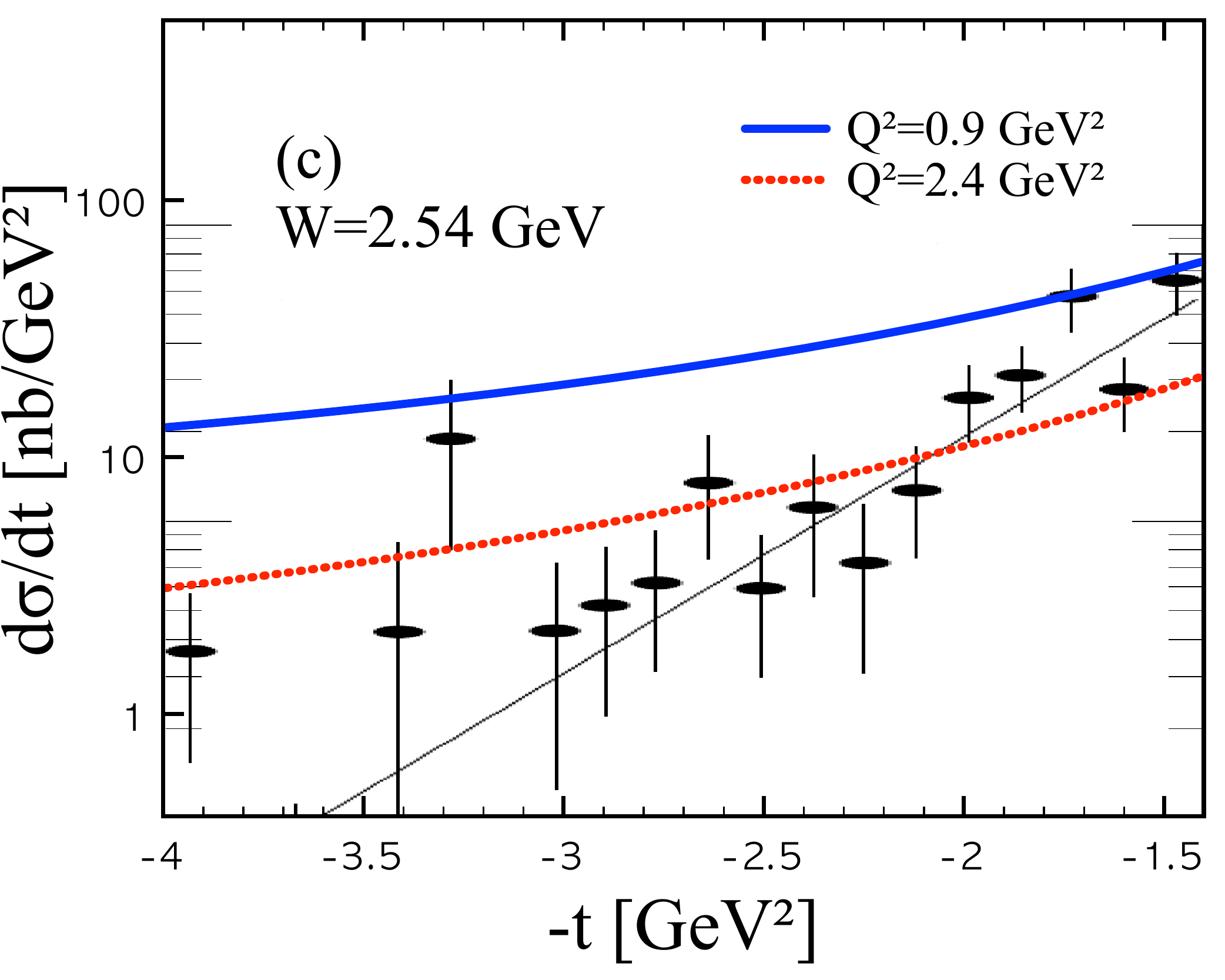}
\end{tabular}
\caption{(Color online) $t$-momentum transfer distribution, $d\sigma/dt$ as functions of $-t$ for $W=2.08$ GeV (a), $2.32$ GeV (b), and $2.54$ GeV (c). The solid and dot lines indicate the experimental photon virtuality interval, $Q^2=0.9\,\mathrm{GeV}^2$ and $2.4\,\mathrm{GeV}^2$, respectively. Experimental data are taken from Ref.~\cite{Barrow:2001ds} and are parameterized by $e^{bt}$ given in the thin solid lines in each panel with $2.3\pm0.1$ (a), $2.4\pm0.2$ (b), and $1.8\pm0.2$ (c).}       
\label{FIG8}
\end{figure}

In Figure~\ref{FIG6}, we plot the $\phi$ distribution in Eq.~(\ref{eq:DF}) as functions of $\cos\phi$ and $\cos\theta$ at $W=2.4$ GeV for $Q^2=1.05\,\mathrm{GeV}^2$ (a), $1.35\,\mathrm{GeV}^2$ (b), $1.65\,\mathrm{GeV}^2$ (c), and $2.10\,\mathrm{GeV}^2$ (d).  As for the electroproduction, for the forward regions of $K^+$, $\cos\theta\gtrsim0.5$, the $\phi$ distribution are dominated by the B-type curves as functions of $\cos\phi$, signaling the dominant $\Lambda^*(S=1/2)$ contribution, being strengthened by the $K$-exchange contribution in the $t$ channel. We also observe the obvious A-type curves for almost all the $\theta$ regions for the photoproduction in the panel (e), in comparison to those for the electroproduction: $\Lambda^*(S=3/2)$ dominates the production process, due to the contact-term contribution. In the panel (a) of Figure~\ref{FIG7}, we draw the $\phi$ distribution for a typical $K^+$ angle, $\theta=30^\circ$, which manifests the difference between the electro and photoproductions. Here, we choose $W=2.4$ GeV and $\sqrt{Q^2}=(0\sim1.45)$ GeV. The difference between the photo (A-type) and electroproduction (B-type) curves are quite obvious from the theoretical results.
%FIGURE>>>
%FIGURE<<<

In Figure~\ref{FIG8}, we draw the numerical results for the $t$-momentum transfer distribution for various intervals $W=2.08$ GeV (a), $2.32$ GeV (b), and $2.54$ GeV (c) for $Q^2=0.9\,\mathrm{GeV}^2$ (solid) and $2.4\,\mathrm{GeV}^2$ (dot). The experimental data are again taken from  Ref.~\cite{Barrow:2001ds}. The thin solid lines denote the parameterized one via $d\sigma/dt\propto e^{bt}$, where $b$ stand for $2.3\pm0.1$ (a), $2.4\pm0.2$ (b), and $1.8\pm0.2$ (c)~\cite{Barrow:2001ds}. Although we are looking at only a single $Q^2$ value for the numerical calculations, the numerical results reproduce the data qualitatively. 
%FIGURE>>>
%FIGURE<<<

\begin{figure}[t]
\begin{tabular}{cc}
\includegraphics[width=8.5cm]{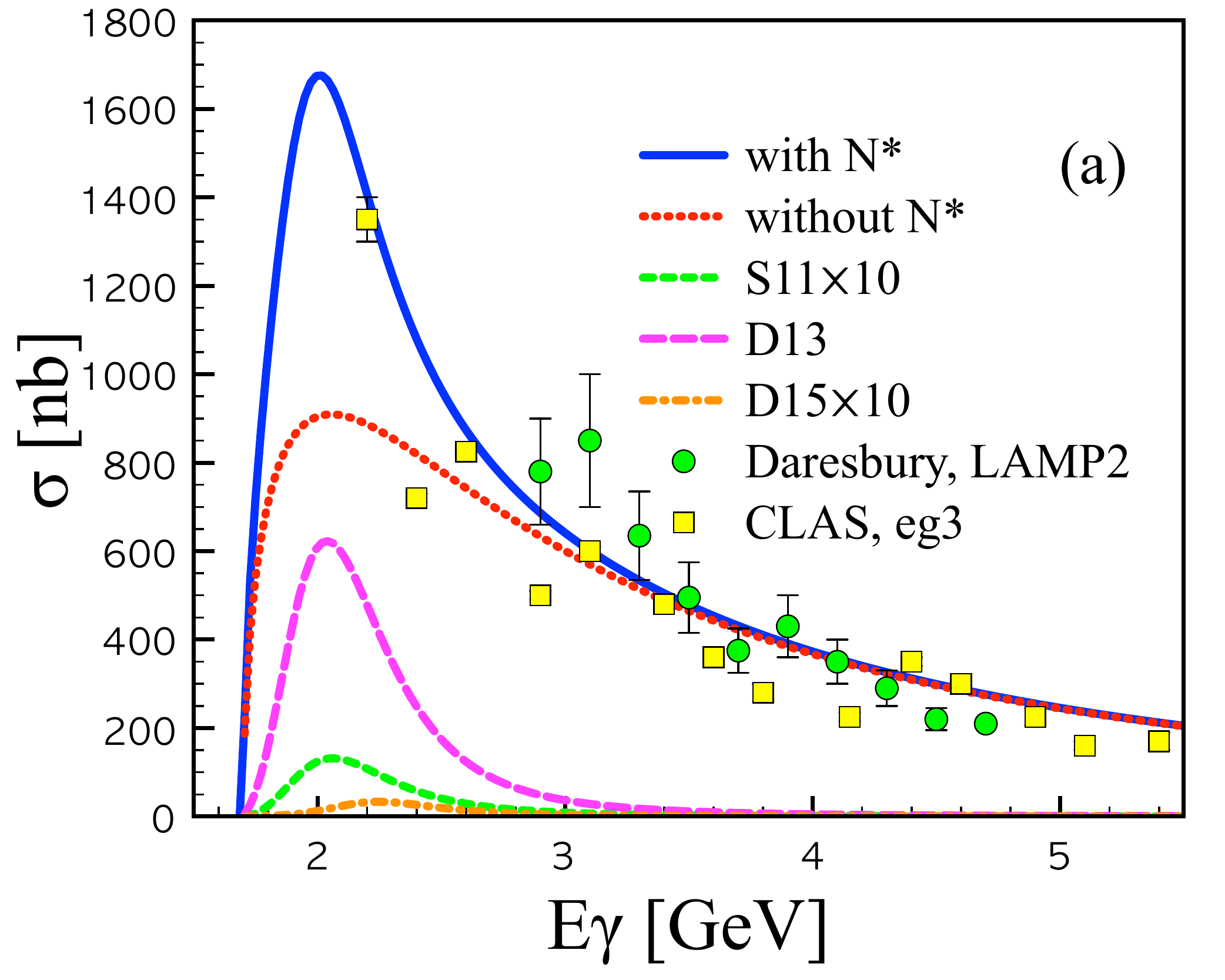}
\includegraphics[width=8.5cm]{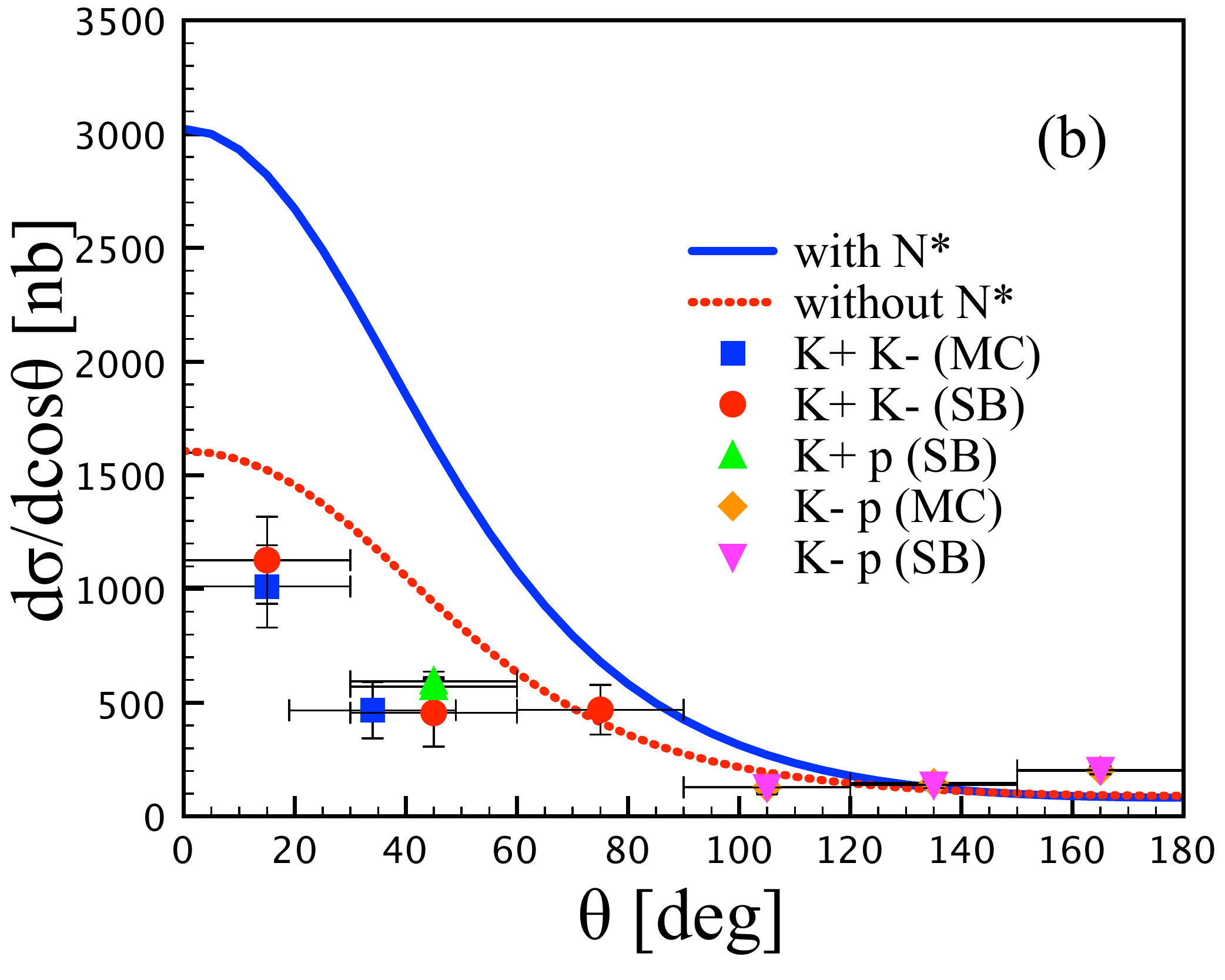}
\end{tabular}
\caption{(Color online) (a) Total cross section for the {\it photoproduction} of $\Lambda(1520)$ ($Q^2=0$ and $\varepsilon=0$) off the proton target with (solid) and without (dot) the resonance contributions. The experimental data are duplicated from the LAMP2 experiment~\cite{Barber:1980zv}. (b) The differential cross sections for the photoproduction are also given as functions of $\cos\theta$ for $E_\gamma=2.15$ GeV in the same manner with the panel (a). The experimental data are taken from Ref.~\cite{Muramatsu:2009zp} with various particle measurement channels using the sideband (SB) and Monte-Carlo (MC) methods.}       
\label{FIG9}
\end{figure}

We now discuss the $\Lambda^*$ photoproduction with the resonance contributions. In the panel (a) of Figure~\ref{FIG9}, we show the numerical results for it off the proton target as functions of $E_\gamma$ with (solid) and without (dot) the resonance contribution, as functions of $E_\gamma$. The circle and square  denote the data from LAMP2~\cite{Barber:1980zv} and eg3-run of CLAS~\cite{Zhao:2010zzm}, respectively. The each resonance contributions are depicted separately. In order to reproduce  data, we choose the parameters as $\Lambda_h=675$ MeV, $\Gamma_{1,2,3}\approx500$ MeV, and $g_{13}=0.63$. We observe that the $D_{13}$ contribution provides considerable enhancement of the production rate in the vicinity of $E_\gamma\approx2$ GeV, reproducing the eg3-run data. The differential cross sections for the photoproduction are also given as functions of $\cos\theta$ for $E_\gamma\approx2.15$ GeV in the panel (b) of Figure~\ref{FIG9} in the same manner with the panel (a). The experimental data are taken from Ref.~\cite{Muramatsu:2009zp} for various meson-meson and meson-baryon channels, using the sideband (SB) and Monte-Carlo (MC) methods. There appears considerable overshoot in the forward scattering region due to the resonance, although the overall shapes of the curves are comparable to the experimental data. This tendency can be related again to the insufficient $Q^2$ dependence for the resonances as shown in the $Q^2$ dependence of the total cross section in the panel (d) of Figure~\ref{FIG4}.  
%FIGURE>>>
%FIGURE<<<

\begin{figure}[t]
\begin{tabular}{cc}
\includegraphics[width=8.5cm]{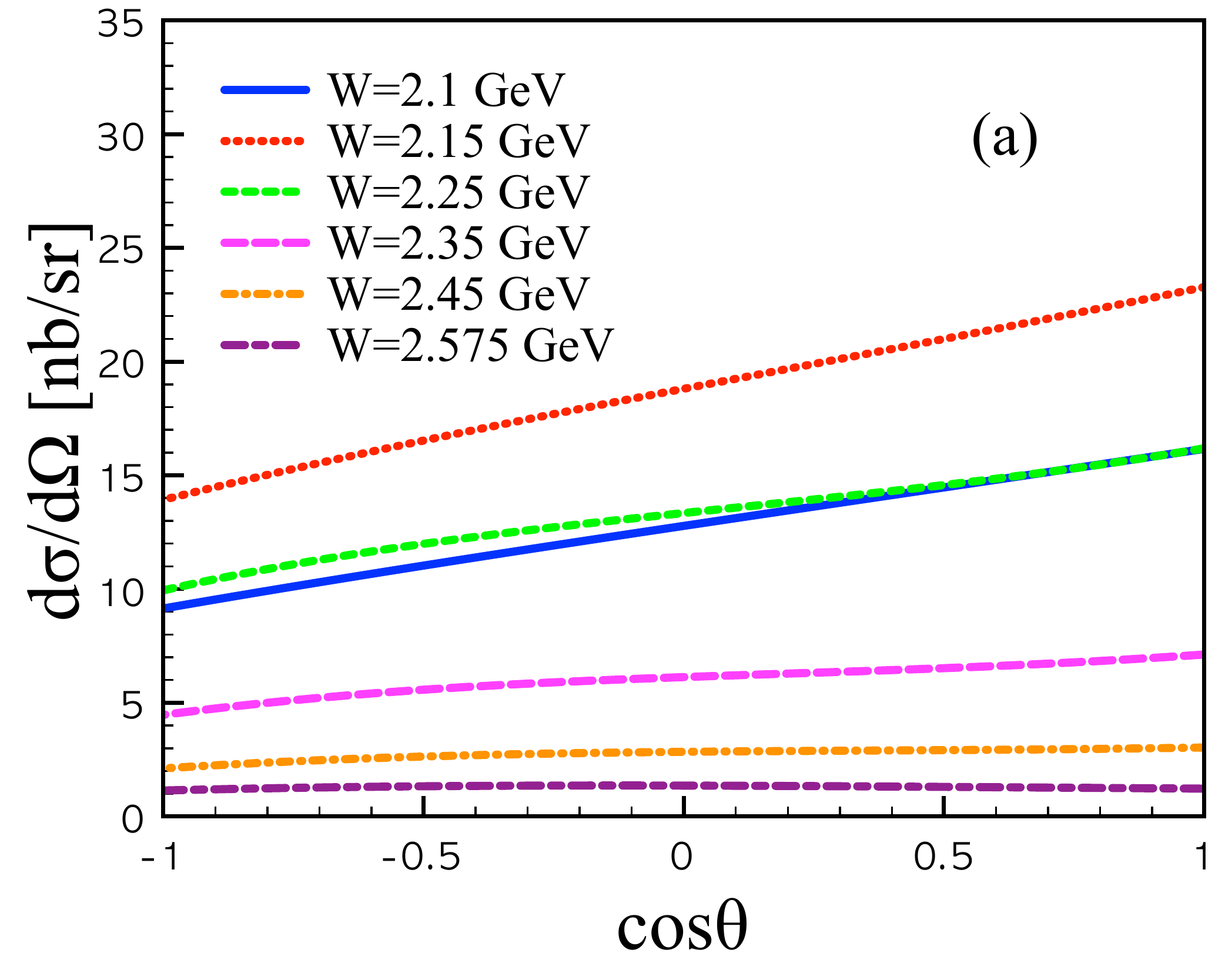}
\includegraphics[width=8.5cm]{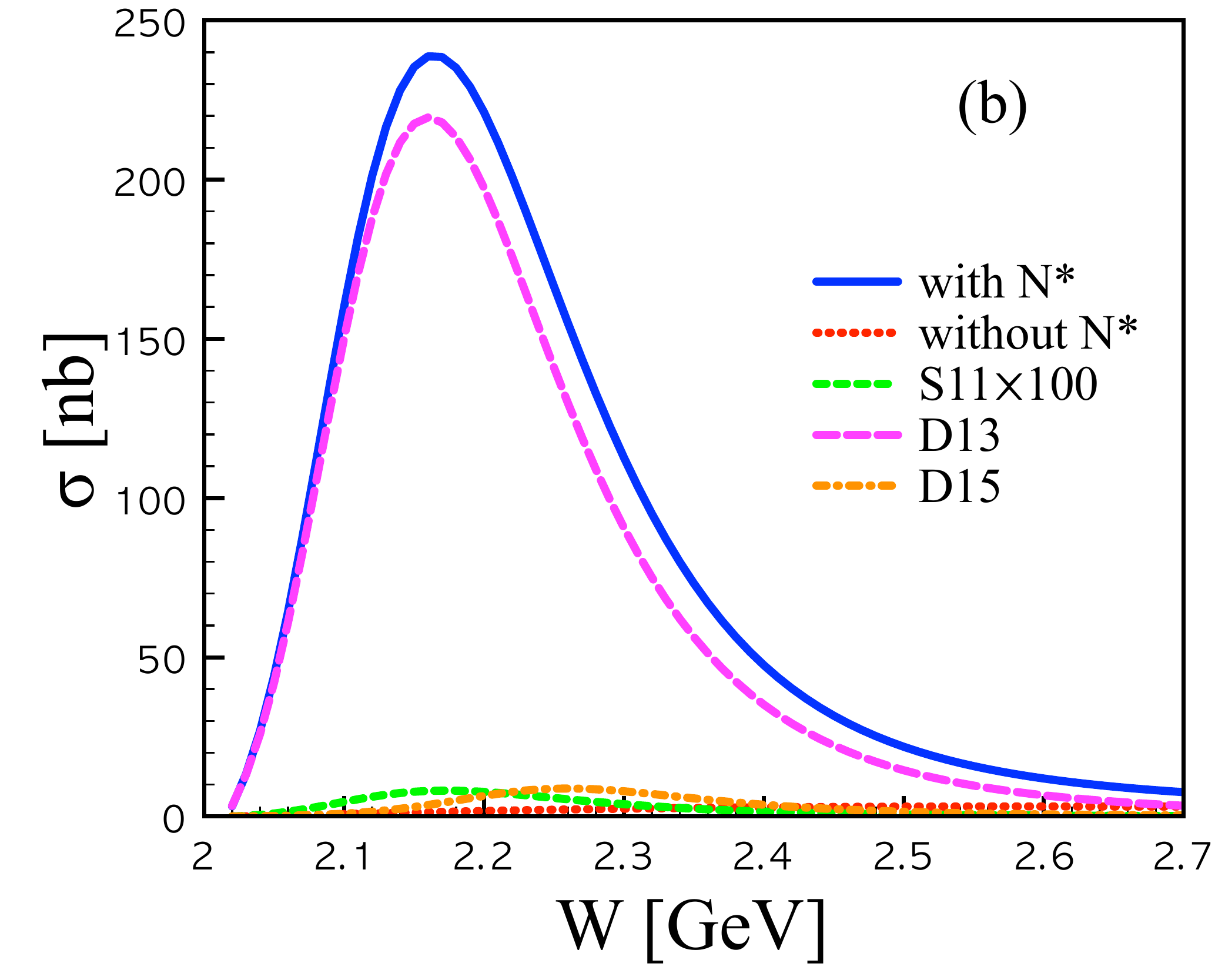}
\end{tabular}
\caption{(Color online) (a) Differential cross sections for the $\gamma^*n\to K^0\Lambda^*$ as functions of $\cos\theta$ for different $W$ values. The $Q^2$ intervals are the same with those given in Figure~\ref{FIG3}. (b) Total cross sections  for the $\gamma^*n\to K^0\Lambda^*$ as functions of $W$  with (solid) and without (dot) the resonance contributions. We also show each resonance contributions. The shaded area indicate the photon virtuality interval $Q^2=(0.9\sim2.4)\,\mathrm{GeV}^2$.}       
\label{FIG10}
\end{figure}

All the parameters, determined for $\gamma^*p\to K^+\Lambda^*$ as above, can be directly used for the neutron-target case, whereas the EM couplings should be changed as in Eq.~(\ref{eq:HR}). In the panel (a) of Figure~\ref{FIG10}, we depict the differential cross sections for $\gamma^*n\to K^0\Lambda^*$ as functions of $\cos\theta$ for different $W$ values, using Eq.~(\ref{eq:HR}). The choices for $W$ are the same with those for each $W$, given in Figure~\ref{FIG3}. We again set $(g_{K^*N\Lambda^*},\kappa_{\Lambda^*})=0$ for simplicity, since we verified that those contributions are almost negligible similar to the proton-target case. If this is the case, the neutron-target production rate is generated almost by the nucleon resonance and $s$-channel magnetic contribution proportional to $\kappa_n$.  We verified that the production rate is almost dominated by the $D_{13}(2150)$ contribution, and the region beyond $W\approx2.35$, in which the resonance effects are diminished, the differential cross sections becomes almost flat due to the $s$-channel nucleon-pole contribution. The total cross sections for the electroproduction off the neutron target as functions of $W$ are given in the panel (b) of Figure~\ref{FIG10} up to $W=2.7$ GeV. As mentioned, the $D_{13}(2150)$ contribution (long-dashed) produces almost all the strength for the total cross section. Hence, we can conclude that the $\Lambda^*$ electroproduction off the neutron target must be a very useful tool to investigate the resonance spectra, due to the negligible background contributions, i.e the absence of the sizable contact-term and $K$-exchange contributions. This conclusion is also valid for the photoproduction with the contact-term dominance.

Now we introduce a quantity, {\it averaged total cross section}, which is defined by 
%EQUATION>>>
\begin{equation}
\label{eq:AVECRO}
\bar{\sigma}=\frac{1}{|W_\mathrm{max}-W_\mathrm{min}|}
\int^{W_\mathrm{max}}_{W_\mathrm{min}}\sigma(W)\,dW.
\end{equation}
%EQUAITON<<<
Employing Eq.~(\ref{eq:AVECRO}), we compute the ratio of the total cross section for the electroproduction off the proton and neutron targets, with and without the resonance contributions, resulting in 
%EQUATION>>>
\begin{equation}
\label{eq:SIGRATIO}
\mathcal{R}^{\gamma^*}_{p/n}(N^*)\equiv\frac{\bar{\sigma}_{\gamma^*p\to K^+\Lambda^*}}{\bar{\sigma}_{\gamma^*n\to K^0\Lambda^*}}\approx\frac{179.01\,\mathrm{nb}}{84.88\,\mathrm{nb}}=2.11,\,\,\,\,
\mathcal{R}^{\gamma^*}_{p/n}(0)\equiv\frac{\bar{\sigma}_{\gamma^*p\to K^+\Lambda^*}}{\bar{\sigma}_{\gamma^*n\to K^0\Lambda^*}}\approx\frac{ 112.06\,\mathrm{nb}}{2.32\,\mathrm{nb}}=48.30,
\end{equation}
%EQUAITON<<<
where we choose $W_\mathrm{(min,max)}=(\mathrm{threshold,2.7})$ GeV for $Q^2=(0.9,\sim2.4)\,\mathrm{GeV}^2$. From the same theoretical calculation with Eq.~(\ref{eq:AVECRO}), we obtain the following values for the photoproduction of $\Lambda^*$ with the resonance contributions,
%EQUATION>>>
\begin{equation}
\label{eq:SIGRATIO2}
\mathcal{R}^{\gamma}_{p/n}(N^*)\equiv\frac{\bar{\sigma}_{\gamma p\to K^+\Lambda^*}}{\bar{\sigma}_{\gamma n\to K^0\Lambda^*}}\approx\frac{ 992.08\,\mathrm{nb}}{ 271.68\,\mathrm{nb}}=3.65,\,\,\,\,
\mathcal{R}^{\gamma}_{p/n}(0)\equiv\frac{\bar{\sigma}_{\gamma p\to K^+\Lambda^*}}{\bar{\sigma}_{\gamma n\to K^0\Lambda^*}}\approx\frac{724.17\,\mathrm{nb}}{ 2.10\,\mathrm{nb}}=344.84.
\end{equation}
%EQUAITON<<<
From these values, ignoring the resonance contributions, we can conclude that the contact-term contribution dominates the photoproduction of $\Lambda^*$ as argued in Refs.~\cite{Nam:2005uq,Toki:2007ab,Nam:2009cv,Nam:2010au}, whereas the contact-term and $K$-exchange contributions are similarly effective for the electroproduction as discussed in the present work. However, this observation is drastically changed by including the resonance contributions as shown above: The neutron- and proton-target cross sections are comparable to each other in their strengths for the electro and photoproductions. We note that this tendency is consistent with the photoproduction experimental data from the eg3-run of CLAS~\cite{Zhao:2010zzm}. It would be very interesting to verify $\mathcal{R}^{\gamma^*}_{p/n}(N^*)$ in Eq.~(\ref{eq:SIGRATIO}) in the future experiment. 
%FIGURE>>>
%FIGURE<<<

%--------------------------------------------------
\section{Summary and outlook}
%--------------------------------------------------
In the present work, we have studied the electroproduction of $\Lambda^*$ off the nucleon target $\gamma^*p\to K^+\Lambda^*$, employing the tree-level Born approximation. Taking into account the results  in our previous work~\cite{Nam:2005uq,Nam:2009cv,Nam:2010au}, we constructed a effective theoretical framework for the present purpose. In addition, $Q^2$ dependences were incorporated with the proton and charged kaon EM form factors. We computed theoretically total and differential cross sections, $\phi$ distribution in the GJ frame, ratio for the $\Lambda^*$ spin states, and $t$-momentum transfer distributions. All the relevant model parameters were determined using the presently available experimental and theoretical information. Below, we list important observations in the present work:
%ITEMIZE>>>
\begin{itemize}
%---------------------
\item The angular and energy dependences for the cross sections for the proton target are reproduced qualitatively well and exhibits the crucial effects from the $D_{13}(2150)$ resonance in the vicinity of $W\approx2.1$ GeV. The resonance enhances the production rate in the forward-scattering region in general, on top of the dominant contact-term and $K$-exchange Born contributions. It also turns out that more realistic $Q^2$ dependence for the resonance contributions would improve the present results, since we employed the proton $Q^2$ dependence even for the resonances for simplicity and the lack of relevant information. 
%---------------------
\item
The longitudinal component $(S_z=0)$ of the virtual photon selects the kaon-exchange contribution in the electroproduction of $\Lambda^*$, being different from the photoproduction case. This selection makes $\Lambda^*(S=1/2)$ increased, being comparable to $\Lambda^*(S=3/2)$, resulting in the different structures of the $\phi$ distribution, as observed in the CLAS electroproduction experiment. We confirm this difference by computing the $\phi$ distribution as functions of $\theta$ and $\phi$ for the electro and photoproductions, separately. From the numerical results, we have the following tendency for the $\Lambda^*$ EM productions off the proton target approximately:
%CENTERING
\begin{center}
$\gamma p\to K^+\Lambda^*$: $\sigma_\mathrm{contact}\gg \sigma_{K}$
\hspace{0.5cm}and\hspace{0.5cm}
$\gamma^*p\to K^+\Lambda^*$: $\sigma_\mathrm{contact}\approx2\,\sigma_K$
\end{center}
%CENTERING
%---------------------
\item
As for the $\Lambda^*$ photoproduction off the proton-target case, the $D_{13}(2150)$ contribution is considerably effective to reproduce the eg3-run data. It also turns out that the resonance effects appear to enhance the forward-scattering production rate. However, this observation can be altered by considering the realistic $Q^2$-dependence for the resonances, as mentioned above.
%---------------------
\item 
The $\Lambda^*$ electroproduction off the neutron target is saturated by the nucleon resonance contributions, such as $D_{13}(2150)$, in the absence of the contact-term and $K$-exchange contributions, due to the electric-charge conservation, as far as the $K^*$-exchange contribution in the $t$ channel is taken to be negligible. From this observation, the $\Lambda^*$ EM production off the neutron target can be considered as one of the best production channels to explore the nucleon-resonance spectra, accompanying with considerably small backgrounds. 
%---------------------
\item
The contact-term dominance, which is one of the key ingredients to understand the background of the $\Lambda^*$ EM production, becomes weak for the electroproduction case according to the resonance effects as well as the enhanced $K$-exchange contribution, although its strength is still sizable in terms of the production rate. This tendency can be also shown alternatively from the ratio of the cross sections off the proton and neutron targets, being averaged from the threshold to $W=2.7$ GeV, with a single resonance contribution from $D_{13}(2150)$:
%CENTERING
\begin{center}
$\mathcal{R}^{\gamma^*}_{p/n}\approx2.11$
\hspace{0.5cm}and\hspace{0.5cm}
$\mathcal{R}^{\gamma}_{p/n}\approx3.65$
\end{center}
%CENTERING
%---------------------
\end{itemize}
%ITEMIZE>>>

As for the next steps to scrutinize the $\Lambda^*$ EM productions, 1) one needs to consider more completed resonance contributions near the threshold region especially for the neutron target case and 2) make clearer the effects from other high-spin strange mesons, including $K^*$. In addition, 3) the realistic $Q^2$ dependence for the resonance will be an important ingredient for the complete studies. Using the present work as a starting point, 4) it is necessary to compute more meaningful physical observables, which manifest the typical production mechanisms of the $\Lambda^*$ EM productions and play the role of useful guides for the future experiments. As for addressing 1), we plan to employ the isobar model~\cite{Drechsel:1998hk} for the $\Lambda^*$ EM productions to include the resonance contributions systematically and to reproduce the neutron data realistically, while the $t$-channel Regge trajectories can be also applied to address 2), taking into account the higher-spin strange meson contributions simultaneously. The Feynman-Regge interpolation prescription, suggested in Ref.~\cite{Nam:2010au}, can be also employed to extend the present results to higher energy regions in a consistent manner. This extension to the higher energy will be useful, considering the planned future upgrade of the beam energy in CLAS of Jefferson laboratory, i.e. CLAS12 at Hall B~\cite{Stepanyan:2010kx}. We are looking for a possibility to employ the EM form factors for the resonances, using an effective low-energy models, such as the quark model~\cite{Aznauryan:2012ec}, to address 3). As for 4), we may compute polarization observables, such as the beam, target, and recoil $\Lambda^*$ polarizations, in addition to the double-polarization ones. Related and combined works are under progress and will appear elsewhere in the near future.
%-------------------------------------------------
\section*{acknowledgment}
%-------------------------------------------------
The author is grateful to A.~Hosaka, C.~W.~Kao, S.~H.~Kim, and S.~H.~Lee for fruitful and stimulating discussions for the present work.
%-------------------------------------------------
\section*{Appendix}
%-------------------------------------------------
Here we discuss about the photon polarization vectors and the definition of the polarization parameter $\varepsilon^*$. The photon polarization vectors in the present work are defined as follows:
%EQUATION>>>
\begin{equation}
\label{eq:PP1}
\epsilon_x=(0,\sqrt{1+\varepsilon^*},0,0),\,\,\,\,
\epsilon_y=(0,0,\sqrt{1-\varepsilon^*},0,0),\,\,\,\,
\epsilon_z=\sqrt{2\varepsilon^*}\left(\frac{k}{|\bm{q}|},0,0,\frac{E_{\gamma^*}}{|\bm{q}|} \right).
\end{equation}
%EQUAITON<<<
and their norms squared are given by
%EQUATION>>>
\begin{equation}
\label{eq:NORM}
\epsilon^2_x=1+\varepsilon^*,\,\,\,\,
\epsilon^2_y=1-\varepsilon^*,\,\,\,\,
\epsilon^2_z=2\varepsilon^*\left(\frac{k^2-E^2_{\gamma^*}}{|\bm{q}|^2} \right)=2\varepsilon^*,
\end{equation}
%EQUAITON<<<
where we have used $k^2=E^2_{\gamma^*}+Q^2$ for the virtual photon.
According to the gauge-invariance argument as in Ref.~\cite{Akerlof:1967zza}, the scattering amplitude can be simplified by omiting the scalar components. This simplification can be realized by setting all the $0$-th components of the photon polarization vectors in Eq.~(\ref{eq:PP1}) to be zero:
%EQUATION>>>
\begin{equation}
\label{eq:PP2}
\epsilon'_x=(0,\sqrt{1+\varepsilon^*}),0,0),\,\,\,\,
\epsilon'_y=(0,0,\sqrt{1-\varepsilon^*}),0,0),\,\,\,\,
\epsilon'_z=\sqrt{2\varepsilon^*}\left(0,0,0,\eta\frac{E_{\gamma^*}}{Q} \right).
\end{equation}
%EQUAITON<<<
Note that we multiply the $4$-th component of $\epsilon'_z$ by $\eta$, which compensates the above simplification. Even with this change, the norm of the polarization vector should not be altered, and this consideration gives the value for $\eta$:
%EQUATION>>>
\begin{equation}
\label{eq:EPP}
\epsilon'^2_z=\epsilon^2_z\to-2\varepsilon^* \eta \frac{E^2_{\gamma^*}}{Q^2}=2\varepsilon^*\to
\eta=-\frac{Q^2}{E^2_{\gamma^*}}.
\end{equation}
%EQUAITON<<<
This multiplication factor $\eta$ is just the same with that given in Ref.~\cite{Akerlof:1967zza}, i.e. $k^2/K^2_0$ in their notation. In literatures, which use the above simplification, motivated by the gauge invariance, an alternative definition $\varepsilon^*_L\equiv|\eta|\varepsilon^*$ is frequently employed for the polarization three vector $\bm{\epsilon}_z=-\sqrt{2\varepsilon^*_L}(0,0,E_{\gamma^*}/Q)$. Note that we have an additional minus sign for the polarization three vector as already argued in Ref.~\cite{Akerlof:1967zza}.

Here, we provide approximated amplitudes for the resonance contributions as an {\it example}. It can be easily verified that, at very low-energy region, the invariant amplitudes for the $D_{13}$ and $D_{15}$ contributions can be written by changing the projection operators with the metric tensors. Substituting $\mathcal{G}^{R_3}_{\mu\nu}\to g_{\mu\nu}$ for $D_{13}$, one is led to
%EQUATION>>>
\begin{eqnarray}
\label{eq:app1}
i\mathcal{M}^{R_3}_{s}&\approx&\frac{e|g_{13}|}{2M_KM_N}\bar{u}_2^\mu\frac{\gamma_5\rlap{/}{k}_3(\rlap{/}{k}_1+\rlap{/}{k}_2+M_N)g_{\mu\nu}}{s-M^2_{R_3}+i\Gamma_{R_3}M_{R_3}}
\left[h_{13}(k_1^\nu\rlap{/}{\epsilon}-\rlap{/}{k}_1\epsilon^{\nu})-\frac{h_{23}}{2M_N}
[k_1^\nu(\epsilon\cdot k_2)-\epsilon^\nu(k_1\cdot k_2)] \right]u_1
\cr
&=&\frac{e|g_{13}|}{2M_KM_N}\bar{u}_2\frac{\gamma_5\rlap{/}{k}_3(\rlap{/}{k}_1+\rlap{/}{k}_2+M_N)}{s-M^2_{R_3}+i\Gamma_{R_3}M_{R_3}}
\left[h_{13}[\rlap{/}{\epsilon}(\varepsilon^*\cdot k_1)-\rlap{/}{k}_1(\varepsilon^*\cdot \epsilon)]
+\frac{h_{23}}{2M_N}[(\varepsilon^*\cdot \epsilon)(k_1\cdot k_2)-(\varepsilon^*\cdot k_1)(\epsilon\cdot k_2)]\right]u_1,
\nonumber
\end{eqnarray}
%EQUAITON<<<
where $\varepsilon$ indicates the polarization vector for $\Lambda^*$. As for $D_{15}$, we choose the following approximation, 
%EQUATION>>>
\begin{equation}
\label{eq:ss}
\mathcal{G}^{R_5}_{\mu\nu\sigma\rho}\to
\frac{1}{2}( g_{\mu\sigma} g_{\nu\rho} 
+g_{\mu\rho} g_{\nu\sigma})
-\frac{1}{5} g_{\mu\nu} g_{\sigma\rho}
-\frac{1}{10}( \gamma_\mu \gamma_\sigma g_{\nu\rho}  
+\gamma_\mu  \gamma_\rho  g_{\nu\sigma}
+\gamma_\nu  \gamma_\sigma g_{\mu\rho} 
+\gamma_\nu  \gamma_\rho  g_{\mu\sigma} ) , 
\end{equation}
%EQUAITON<<<
then we have
%EQUATION>>>
\begin{eqnarray}
\label{eq:SAS}
i\mathcal{M}^{R_5}_{s}&\approx&\frac{e|g_{15}|e^{i\phi_\mathrm{phase}}}{4M^2_KM^2_N}\bar{u}_2^\mu\frac{\rlap{/}{k}_3(\rlap{/}{k}_1+\rlap{/}{k}_2+M_N)}{s-M^2_{R_5}+i\Gamma_{R_5}M_{R_5}}\left[\mathrm{M}_1-\mathrm{M}_2 \right]
 u_1F_R(s),
\end{eqnarray}
%EQUAITON<<<
where we have used the following notations for convenience:
%EQUATION>>>
\begin{eqnarray}
\label{eq:SASD}
&&\mathrm{M}_1=\left[h_{15}\rlap{/}{\epsilon}-\frac{h_{25}(\epsilon\cdot k_2)}{2M_N} \right]
\cr
&&\times\left[\frac{(k_1\cdot k_3)(\varepsilon^*\cdot k_1)+(k_1\cdot k_3)(\varepsilon^*\cdot k_1)}{2}-\frac{(k_1\cdot k_1)(\varepsilon^*\cdot k_3)}{5}
-\frac{(\varepsilon^*\cdot k_1)\rlap{/}{k}_3\rlap{/}{k}_1
+(\varepsilon^*\cdot k_1)\rlap{/}{k}_3\rlap{/}{k}_1
+(k_1\cdot k_3)\rlap{/}{\varepsilon}^*\rlap{/}{k}_1
+(k_1\cdot k_3)\rlap{/}{\varepsilon}^*\rlap{/}{k}_1}{10} \right],
\cr
&&\mathrm{M}_2=\left[h_{15}\rlap{/}{k}_1-\frac{h_{25}(k_1\cdot k_2)}{2M_N} \right]
\cr
&&\times\left[\frac{(k_1\cdot k_3)(\varepsilon^*\cdot\epsilon)+(\epsilon\cdot k_3)(\varepsilon^*\cdot k_1)}{2}-\frac{(\epsilon\cdot k_1)(\varepsilon^*\cdot k_3)}{5}
-\frac{(\varepsilon^*\cdot k_1)\rlap{/}{k}_3\rlap{/}{\epsilon}
+(\varepsilon^*\cdot \epsilon)\rlap{/}{k}_3\rlap{/}{k}_1
+(k_1\cdot k_3)\rlap{/}{\varepsilon}^*\rlap{/}{\epsilon}
+(\epsilon\cdot k_3)\rlap{/}{\varepsilon}^*\rlap{/}{k}_1}{10}\right].
\end{eqnarray}
%EQUAITON<<<
%-------------------------------------------------

%--------------------------------------------------
%--------------------------------------------------
%--------------------------------------------------
\end{document}